\documentclass[useAMS,usenatbib]{mn2e}
\pdfoutput=1
\usepackage{graphicx}
\usepackage{amssymb}
\usepackage{longtable}

\newcommand{\um}{$\mu$m}
\newcommand{\kms}{km s$^{-1}$}
\newcommand{\cmdue}{cm$^{-2}$}
\newcommand{\cmtre}{cm$^{-3}$}

\newcommand{\msun}{M$_{\odot}$}
\newcommand{\ncol}{$N_{\rm H_2}$}

\title[Filaments in the Lupus molecular clouds]{Filaments in the Lupus molecular clouds}

\author[M. Benedettini et al.]{M. Benedettini$^{1}$\thanks{E-mail: milena.benedettini@inaf.it}, E. Schisano$^{1}$, S. Pezzuto$^{1}$, D. Elia$^{1}$, P. Andr\'{e}$^{2}$, V. K\"{o}nyves$^{2}$, \newauthor
N. Schneider$^{3}$,  P. Tremblin$^{4,5}$, D. Arzoumanian$^{2}$, A.M. di Giorgio$^{1}$, J. Di Francesco$^{6,7}$, \newauthor T. Hill$^{2}$, S. Molinari$^{1}$, 
F. Motte$^{2}$, Q. Nguyen-Luong$^{8,9}$, P. Palmeirim$^{2}$, \newauthor A. Rivera-Ingraham$^{10,11}$, A. Roy$^{2}$, K.L.J. Rygl$^{12}$, 
L. Spinoglio$^{1}$, D. Ward-Thompson$^{13}$,  \newauthor G. J. White$^{14,15}$\\
$^{1}$INAF -- Istituto di Astrofisica e Planetologia Spaziali, via Fosso del Cavaliere 100, 00133 Roma\\
$^{2}$ Laboratoire AIM, CEA/DSM-CNRS-Universit\'{e} Paris Diderot, IRFU/Service d'Astrophysique, Saclay, Orme des Merisiers, 91191 Gif\\
$^{3}$ Universit\'{e} Bordeaux, LAB/OASU, CNRS, UMR 5804, 33270 Floirac\\
$^{4}$ Astrophysics Group, University of Exeter, EX4 4QL Exeter\\
$^{5}$ Maison de la Simulation, CEA-CNRS-INRIA-UPS-UVSQ, USR 3441, Saclay, 91191 Gif-Sur-Yvette\\
$^{6}$ Department of Physics \& Astronomy, University of Victoria, BC, V8W 3P6, Victoria\\
$^{7}$ National Research Council Canada, 5071 West Saanich Road, BC, V9E 2E7, Victoria\\
$^{8}$ National Astronomical Observatory of Japan, Chile Observatory, 2-21-1 Osawa, Mitaka, Tokyo 181-8588\\
$^{9}$ Canadian Institute for Theoretical Astrophysics, University of Toronto, 60 St. George Street, ON M5S 3H8, Toronto\\
$^{10}$ Universit\'{e} de Toulouse, UPS-OMP, IRAP, 31028, Toulouse \\
$^{11}$ CNRS, IRAP, 9 Av. Colonel Roche, BP 44346, 31028, Toulouse cedex 4\\
$^{12}$ INAF -- Istituto di Radioastronomia, Via Gobetti 101, 40129, Bologna \\
$^{13}$ Jeremiah Horrocks Institute, University of Central Lancashire, Preston, PR1 2HE\\
$^{14}$ The Rutherford Appleton Laboratory, Chilton, Didcot, OX11 0NL\\
$^{15}$ Department of Physics and Astronomy, The Open University, Milton Keynes\\
}
\begin{document}

\date{Accepted 2015 July 29.  Received 2015 July 29; in original form 2015 April 21. }

\pagerange{\pageref{firstpage}--\pageref{lastpage}} \pubyear{2013}

\maketitle

\label{firstpage}

\begin{abstract}
%abstract, normally of not more than 200 words 

We have studied the filaments extracted from the column density maps of the nearby Lupus 1, 3, and 4 molecular clouds, derived from photometric maps observed with the {\it Herschel} satellite. Filaments in the Lupus clouds have quite low column densities, with a median value of $\sim$1.5$\times$10$^{21}$ \cmdue\, and most have masses per unit length lower than the maximum critical value for radial gravitational collapse. Indeed, no evidence of filament contraction has been seen in the gas kinematics. We find that some filaments, that on average are thermally subcritical, contain dense cores that may eventually form stars. This is an indication that in the low column density regime, the critical condition for the formation of stars may be reached only locally and this condition is not a global property of the filament. Finally, in Lupus we find multiple observational evidences of the key role that the magnetic field plays in forming filaments, and determining their confinement and dynamical evolution.

\end{abstract}

\begin{keywords}
ISM: clouds -- stars: formation.
\end{keywords}

\section{Introduction}

Filaments are ubiquitous structures present in both low- and in high-mass star-forming regions (e.g., \citealt{hatchell05}; \citealt{goldsmith08}; \citealt{andre10}; \citealt{molinari10}). They were recognised more than 30 years ago \citep{schneider79} but only recently the {\it Herschel}\footnote{{\it Herschel} is an ESA space observatory with science instruments provided by European-led Principal Investigator consortia and with important participation from NASA.} Space Observatory has shown the omnipresence of the filamentary structures throughout star-forming regions and their strict association with the dense cores from which new stars will form (e.g., \citealt{andre10}; \citealt{molinari10}; \citealt{polychroni13}). The {\it Herschel} results have focused special attention on the formation mechanism of filaments, and in understanding their role in the star formation process, with several studies either analysing interesting filaments, or adopting a more statistical approach. These studies, although far from answering the key questions, have started to reveal important properties. For example, filaments in molecular clouds are characterised by having  a narrow distribution of widths, with a median value of 0.10$\pm$0.03 pc \citep{arzoumanian11}. This corresponds, within a factor of ∼2, to the sonic scale below which interstellar turbulence becomes subsonic in diffuse gas. This similarity in scale supports the argument that filaments may form as a result of the dissipation of large-scale turbulence. Lower column density filaments are thermally subcritical and gravitationally unbound, and are probably confined by external pressure, while higher column density filaments are thermally supercritical and gravitationally bound and show evidences of gravitational collapse (e.g., \citealt{schisano14}; \citealt{arzoumanian13}; \citealt{nguyen13}; \citealt{hill12}). A kinematical study conducted on the B211/213 filament in Taurus \citep{hacar13}, revealed that some filaments are actually composed of several distinct, coherent velocity components of gas. In the more massive filaments, spectral observations indicate a more dynamical condition of the gas, revealing velocity gradients indicative of collapse and contraction (\citealt{schneider10}; \citealt{kirk13}; \citealt{peretto14}).

The Lupus complex is an interesting region in which to study filaments. It is one of the nearest low-mass star-forming regions (distance between 150 pc and 200 pc, \citealt{comeron08}) and a few long and well defined filaments have been identified in it (\citealt{rygl13}: \citealt{benedettini12};  \citealt{poidevin14}; \citealt{matthews14}). The complex is located in the Scorpius--Centaurus OB association, whose massive stars are likely to have played a significant role in shaping the diffuse material and in the general evolution of the complex. 

In this paper, we present an analysis of filaments extracted from the column density maps of the Lupus 1, 3, and 4 star-forming regions, produced from the photometric observations carried out with the {\it Herschel} satellite. 
In Sect. 2 we describe the observations and the data reduction. The analysis of the filament properties and of the probability distribution function is presented in Sect. 3 ad 4, respectively. The results are discussed in Sect. 5 and the main conclusions are reported in Sect. 6.

\section{Observations and data reduction}

\subsection{{\it Herschel} photometric maps}

As part of the {\it Herschel} Gould Belt Survey (HGBS\footnote{http://gouldbelt-herschel.cea.fr}; \citealt{andre10}) the three subregions of the Lupus complex, Lupus 1, 3, and 4, have been observed in five photometric bands between 70 \um\, and 500 $\mu$m with the Photodetector Array Camera and Spectrometer (PACS; \citealt{poglitsch10}) and the Spectral and Photometric Imaging Receiver (SPIRE; \citealt{griffin10}) on-board the {\it Herschel} Space Observatory \citep{pilbratt10}. The observations were carried out in the parallel observing mode with a scanning velocity of 60 arcsec/sec and each zone was observed twice with orthogonal scanning directions. 
The spatial resolution in this observing mode is 7.6\arcsec\, at 70 \um, 11.5\arcsec\, at 160 \um,  18\farcs2 at 250\um, 24\farcs9 at 350 \um\, and 36\farcs3 at 500 \um.
A first version of the Lupus maps was published by \cite{rygl13} who used the {\sc romagal} pipeline \citep{traficante11}. In this paper we present a new reduction of the maps obtained with a different pipeline and updated calibration files. The pipeline for the production of the PACS maps (at 70 \um\, and 160 \um) consisted of the following steps: the Level 1 data products were generated with Standard Product Generation (SPG) version 10 within the {\it Herschel} Interactive Processing Environment ({\sc hipe}\footnote{{\sc hipe} is a joint development by the {\it Herschel} Science Ground Segment Consortium, consisting of ESA, The NASA {\it Herschel} Science Center, and the HIFI, PACS and SPIRE consortia}); Level 1 data were arranged in time ordered data (TOD) and exported in fits format; the TODs of the two orthogonal observations were ingested into the {\sc unimap} map maker \citep{piazzo15} to produce the final maps. The SPIRE data (at 250 \um, 350 \um, and 500 \um) were reduced using modified pipeline scripts of {\sc hipe} version 10. Data taken during the turnaround of the satellite were included in order to assure a larger area covered by the observations and enough overlap. Because the observed Galactic regions are dominated by extended emission of the interstellar medium, extended emission relative gain factors were applied to the bolometer time-lines. The resulting Level 1 contexts for each scan direction were then combined using the naive map maker in the destriper module. 

The final absolute flux calibration for both PACS and SPIRE maps was performed by adding a zero-level offset derived by comparing the {\it Herschel} data with the {\it Planck} and {\it IRAS} data of the same area of the sky. A dust model has been assumed for extrapolating the flux at the {\it Herschel} wavelengths \citep{bernard10}. The final flux accuracy is better than 20 per cent. Notably, the Lupus 1 maps were affected by stray Moonlight, visible as a bright vertical band in each image. We removed this contamination by evaluating the stray-light contribution from the difference between the observed map (affected by stray-light) and the model used for the flux calibration. This procedure worked well but not for the 70 \um\, image for which the correction could not be applied. The 70 \um\, map, however, was not used for the production of the column density maps (see Sect. \ref{sec_cdmap}), that therefore are corrected from the stray-light effect.

\subsection{CS (2--1) line maps}

A CS (2--1) line map of the Lupus 1 and 3 clouds was acquired at the 22-m {\it Mopra} radio telescope using the On The Fly observing mode with the narrow band mode of the UNSW-Mopra Spectrometer (UNSW-MOPS) digital filterbank back-end. The velocity resolution at 97.981 GHz is 0.1 \kms\, the typical rms of the single spectrum is $\sim$ 0.4 K and the Half Power Beam Width is 35\ arcsec. Part of these maps have been already published in \citet{benedettini12}, where details regarding the data reduction can be found. In May 2012, we acquired new CS (2--1) spectra toward the Lupus 3 cloud extending the coverage of the region with respect to the previous published map. The CS (2--1) maps cover only the brighter and denser part of the Lupus 1 and 3 clouds, approximately the region above the 4 mag of visual extinction.

\subsection{Column density maps}
\label{sec_cdmap}

\begin{figure*}
\includegraphics[scale=.8]{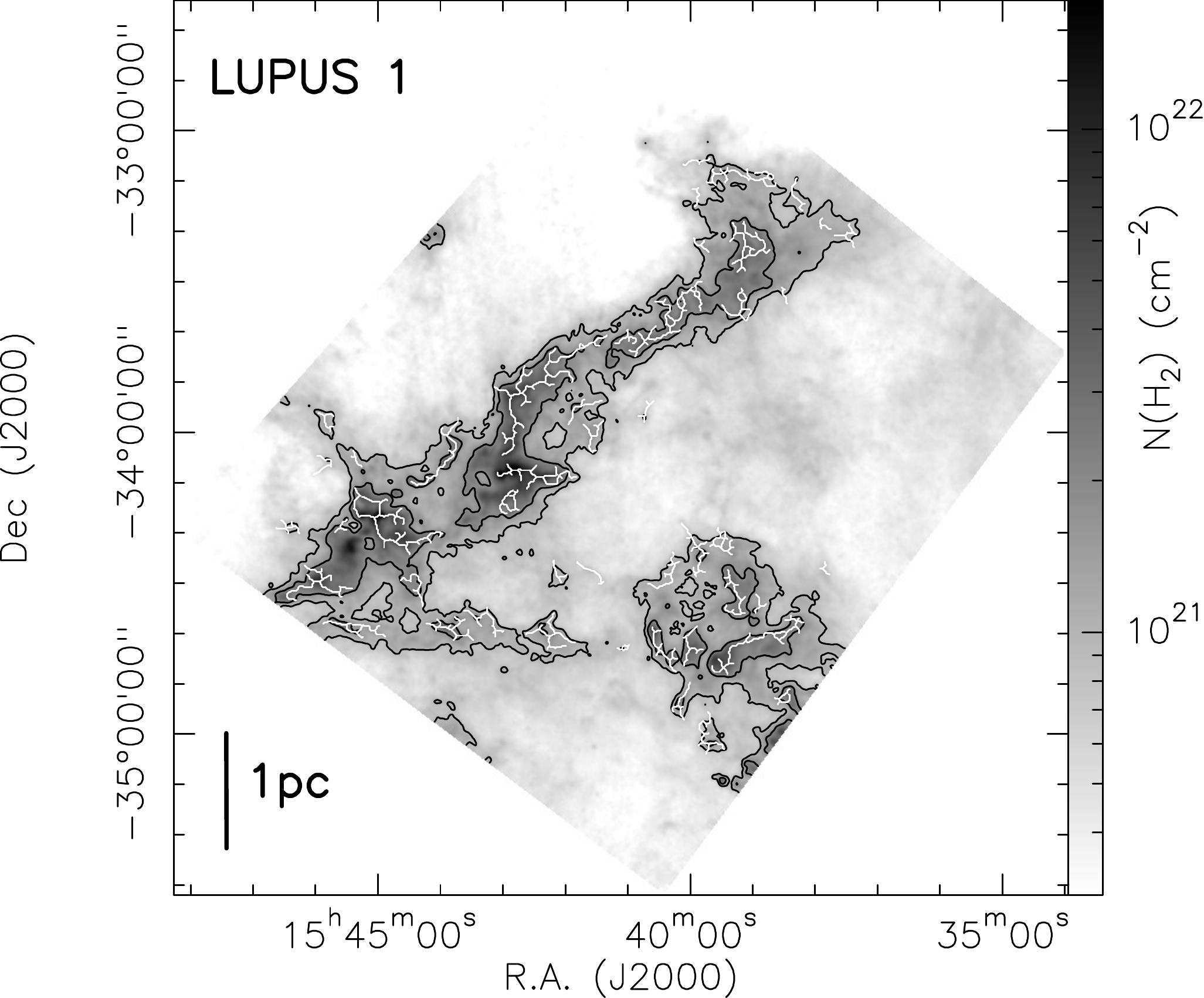}
\caption{Column density map at 36 arcmin resolution derived form {\it Herschel}  data of the Lupus 1 cloud. The black contours are $A_{\rm v}$ = 1 mag and 2 mag, respectively. The white lines identify the spines of all the branches of the filaments.}
\label{fig:1_lup1}       
\end{figure*}

\begin{figure*}
\includegraphics[scale=.8]{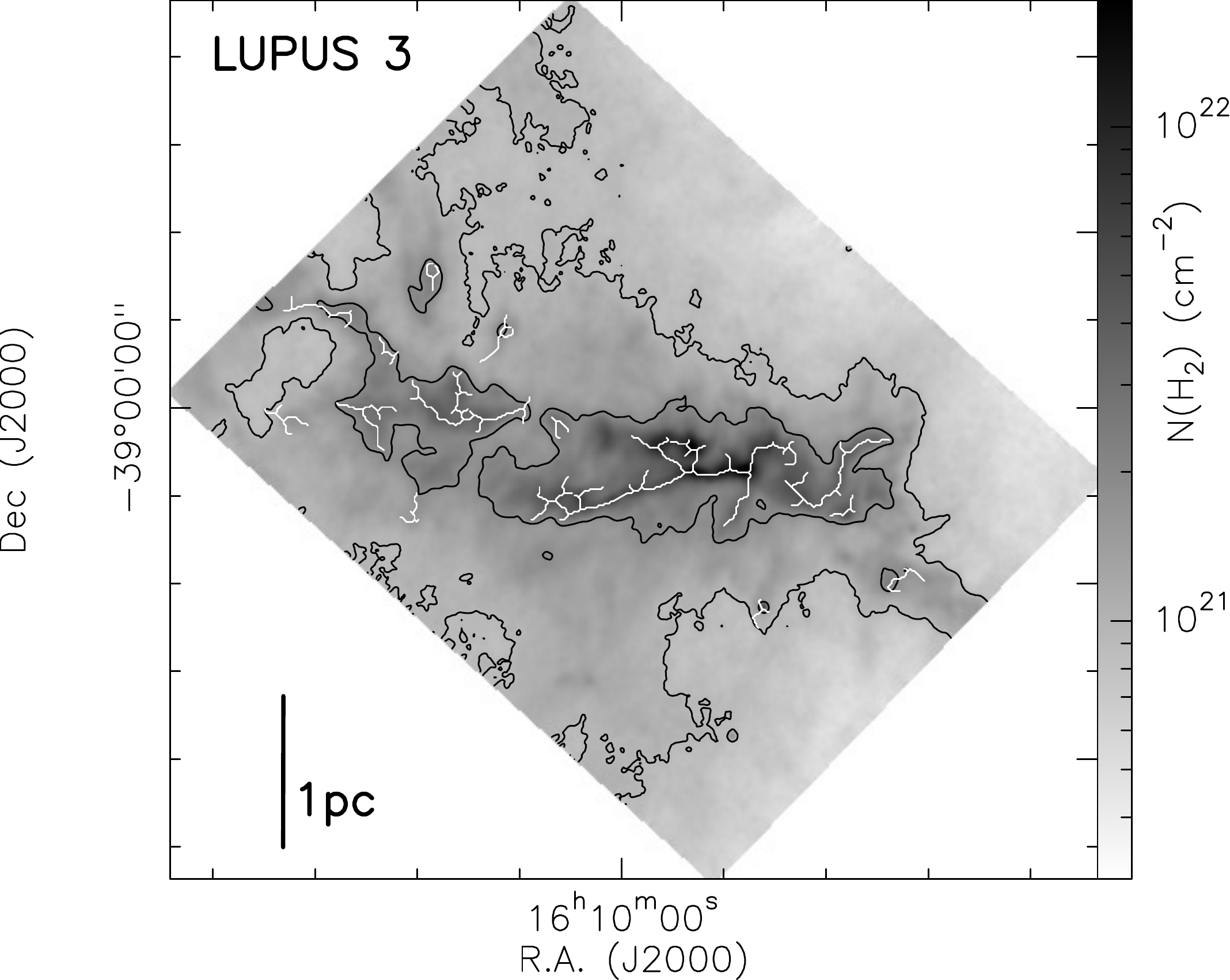}
\caption{As Fig. \ref{fig:1_lup1} for Lupus 3.}
\label{fig:1_lup3}       
\end{figure*}

\begin{figure*}
\includegraphics[scale=.58]{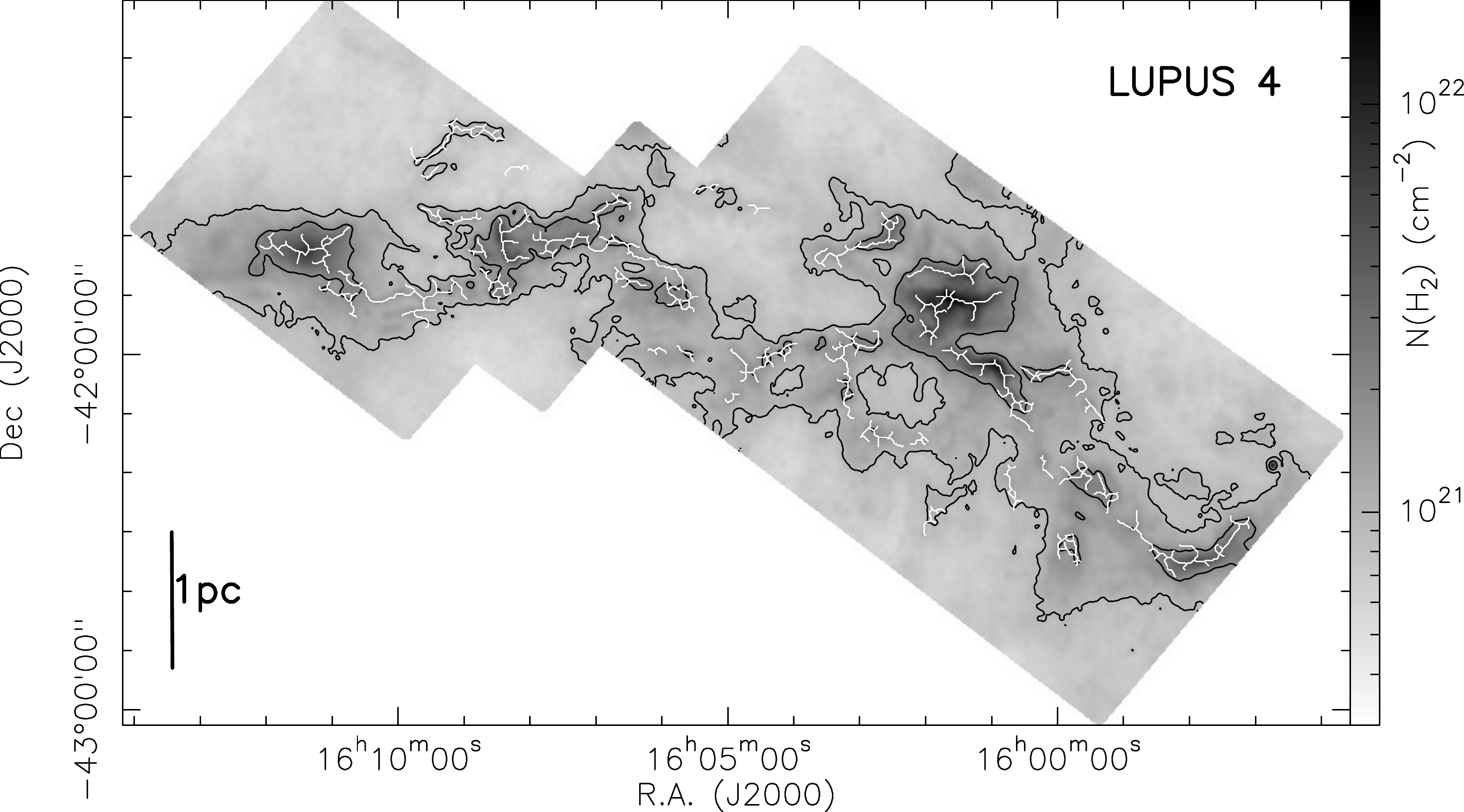}
\caption{As Fig. \ref{fig:1_lup1} for Lupus 4.}
\label{fig:1_lup4}       
\end{figure*}

The column density maps of the three Lupus clouds are shown in Fig. \ref{fig:1_lup1}, \ref{fig:1_lup3} and \ref{fig:1_lup4}. The maps were produced by applying a grey-body fit in each pixel to the Spectral Energy Distribution (SED) of the four surface brightness values at 160 \um, 250 \um, 350  \um\, and 500 \um\, smoothed to a common spatial resolution of 36 arcsec, i.e. the resolution of the 500 \um\, map and resampled with a pixel size of 14 arcsec. 
The spatial resolution of the column density maps corresponds to a linear scale of 0.026 pc for Lupus 1 and 4 (distance of 150 pc) and 0.035 pc for Lupus 3 (distance of 200 pc).
For the grey-body fitting, we assumed a dust opacity of $\kappa_{\rm 250}$=0.144 cm$^2$g$^{-1}$ (already accounting for a gas-to-dust ratio of 100), a grain emissivity parameter $\beta$=2 (cf. Hildebrand 1983) and a mean molecular weight $\mu$=2.8, values adopted as the standard by the HGBS consortium. Previous works (e.g., \citealt{roy14}, \citealt{sadavoy13}) estimated that the column densities derived with these assumptions are accurate to 50 per cent for dense cores. The H$_2$ column density towards the Lupus clouds is typically low, with the denser region corresponding to visual extinction between 3 mag and 6 mag and a very small portion of the clouds at $A_{\rm v} >$ 7 mag.

\begin{table}
 \centering
  \caption{Parameters of the Lupus clouds.}
  \begin{tabular}{@{}lccc@{}}
  \hline                                   & Lupus 1 & Lupus 3 & Lupus 4 \\
Distance (pc)                              & 150     & 200     & 150 \\
area above $A_{\rm V}>$2 mag (deg$^2$)     & 0.32    & 0.19    & 0.22 \\
Mass above $A_{\rm V}>$2 mag (\msun)       & 163     & 162     & 106 \\
%Mass in filaments (\msun)                  & 205     & 147     & 166 \\
\hline 
\end{tabular}
\label{param}
\end{table}

We estimated the total mass of the clouds by summing the mass inside the $A_{\rm v}$=2 mag contour, using the conversion formula \ncol = 9.4$\times$10$^{20}$ $A_{\rm v}$ \citep{bohlin78} and the derived values are listed in Table \ref{param}. We obtained masses lower than the estimates previously quoted by \citet{rygl13} from the {\it Herschel} maps, namely 830 \msun\, (the map was not corrected for the stray-light), 570 \msun\, and 500 \msun\, for Lupus 1, 3 and 4, respectively. The discrepancy is mainly due to the fact that they integrated over an area larger than our area. In fact, the original five bands maps from which \citet{rygl13} derived their column density maps were produced with a different map-maker and older calibration files and they used a different opacity law.

\begin{figure*}
\includegraphics[angle=90,scale=.78]{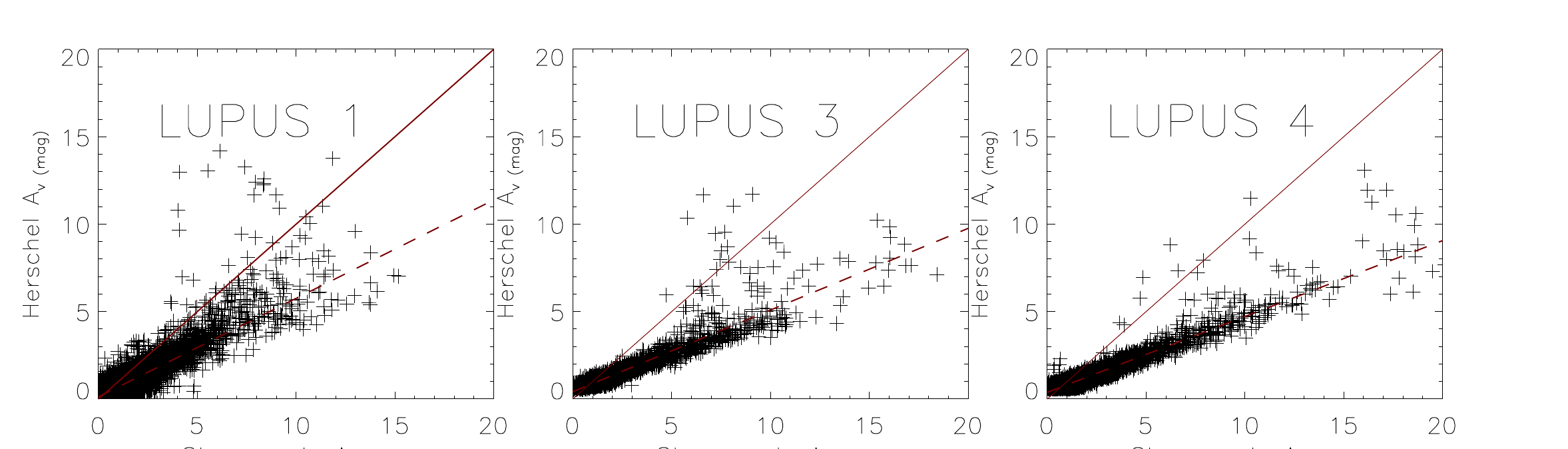}
\caption{Comparison between the visual extinction derived from the {\it Herschel} maps and the star counts maps derived from NIR data following the method of \citet{cambresy99}. The continuous line identifies the correspondence between the two values, the dashed line is the linear fit to the data. The {\it Herschel} based values are systematically lower than the values derived from star counts.}
\label{fig_av_comp}       
\end{figure*}

In the literature, we have found a variety of estimates for the mass of the Lupus clouds, derived by using extinction or molecular line maps, spanning a large range of values.
A direct comparison between our mass estimates and the literature values is not so trivial since they are derived from maps covering different area of the clouds; moreover the star counts extinction maps and the CO maps have spatial resolution of a few arcminutes and are usually spatially undersampled. Generally speaking, our masses are more similar to the values derived from $^{13}$CO (2--1) line maps \citep{tothill09} and are lower than previous estimates from other extinction maps (\citealt{merin08}; \citealt{cambresy99}). In order to do a more accurate comparison we produced new visual extinction maps by using the star count method of \citet{cambresy99} applied to near-infrared (NIR) data. These new extinction maps cover the same region mapped with {\it Herschel} and have a spatial resolution of 2 arcmin. In Fig. \ref{fig_av_comp} we compare the visual extinction derived from the NIR data with that derived from {\it Herschel} data, in all the common spatial pixels, smoothed at the common resolution of 2 arcmin. We find a good correlation between the two values, but above the threshold of $\sim$ 2 mag the {\it Herschel} values are lower of a factor of $\sim$ 2 than the star counts values, although the higher spatial resolution of the original {\it Herschel} maps. Similarly, column density higher than the {\it Herschel} values have been derived from {\it Planck} data at the even coarse spatial resolution of 10 arcmin \citep{planck15}. It is worth noting that the {\it Herschel} maps detect the dust emission and the extinction maps derived from these measurements require the assumption of the dust opacity, the grain emissivity and the dust to gas ratio. It is therefore possible that the observed discrepancy arises from a wrong assumption in the dust parameters that we used in deriving the column density maps. Values of visual extinction more similar to the star counts values could be reproduced from the {\it Herschel} data halving the opacity at the reference frequency, i.e. assuming $\kappa_{\rm 250}$=0.072 cm$^2$g$^{-1}$. 
In fact, several studies demonstrated that the dust opacity is not constant in the diffuse ISM but it is correlated to the Hydrogen column density and more in general to the volume density. For example, \citet{planck11a} found $\kappa_{\rm 250}$= (0.026--0.069) cm$^2$g$^{-1}$ for high latitude diffuse ISM where the gas is mainly in atomic form, while in the Taurus molecular cloud \citet{planck11b} found $\kappa_{\rm 250}$=0.049 cm$^2$g$^{-1}$ for the atomic phase gas and $\kappa_{\rm 250}$=0.099 cm$^2$g$^{-1}$ for the molecular gas. \citet{martin12} found $\kappa_{\rm 250}$=(0.086 -- 0.17) cm$^2$g$^{-1}$ for a sample of 14 regions near the Galactic plane toward the Vela molecular cloud, mostly selected to avoid regions of high column density ($N(\rm H) >$ 10$^{22}$ \cmdue). Finally, \citet{roy13} analysing the {\it Herschel} data of the Orion A molecular cloud found that the dust opacity in the region increases from 0.043 to 0.129 cm$^2$g$^{-1}$ as function of $N(\rm H)^{0.28}$ over the range of $N(\rm H)$ from 1.5 $\times$10$^{21}$ \cmdue\, to 50 $\times$ 10$^{21}$ \cmdue. Looking at these results, the dust opacity of $\kappa_{\rm 250}$=0.144 cm$^2$g$^{-1}$ assumed in the HGBS consortium for the SED fitting of the dense cores could be too high for the calculation of the column density map in the Lupus complex, which column density is typically lower than the values observed towards dense cores. However, for consistency with the other studies of the HGBS project, in the following analysis we assumed the HGBS standard, bearing in mind that the derived column density values could be underestimated by a factor of two.

%The {\it Herschel} maps, therefore, trace well the molecular gas component, and indeed we find a good agreement between the {\it Herschel} column density maps and the $^{13}$CO (2--1) line maps \citep{tothill09} both in term of emitting area and derived mass. On the other hand, visual extinction maps derived from star counts trace both the molecular and the atomic gas and the higher values of these maps respect to the {\it Herschel} values suggests that in the Lupus clouds about half of the gas could be in the form of atomic Hydrogen, not well traced by {\it Herschel}. The possible presence of a  significant amount of Hydrogen not yet converted in molecular form is an indication of the youth of these regions.

\subsection{Filament detection}
\label{par_fildet}

As {\it Herschel} has largely shown, molecular clouds are mainly structured in filaments and we see such structures also in the Lupus clouds. In order to characterise the filamentary structure of the Lupus clouds, we applied an algorithm that uses the eigenvalues of the Hessian matrix of the column density map to identify the regions belonging to filaments. At the first order, the border of the filament is identified as the place with the maximum variation of the contrast of the column density field. Doing this, the pixels selected will always belong to regions where the curvature of the brightness profile in the maximum curvature direction is within the convexity region. This implies in general a conservative identification of the filament region, because it would approximate the contribution of the wings of the filaments since we do not include the region where the emission profile changes concavity before joining the background emission. In order to recover, at least part of, the external wings of the filaments we applied a morphological dilation operator at the first order border identified with the thresholding of the curvature matrix. The spine of the filament is then derived by applying a morphological operator to thin the pixels inside the borders, connected with a Minimum Spanning Tree (MST) technique. Because of the interwined nature of the filamentary structures, each filamentary region is composed by one or multiple spines connected in nodal points; we call branch a single portion of a filament connecting two nodes. A detailed description of the filament detection algorithm and its performance can be found in \cite{schisano14}. 

\begin{figure}
\includegraphics[scale=.45]{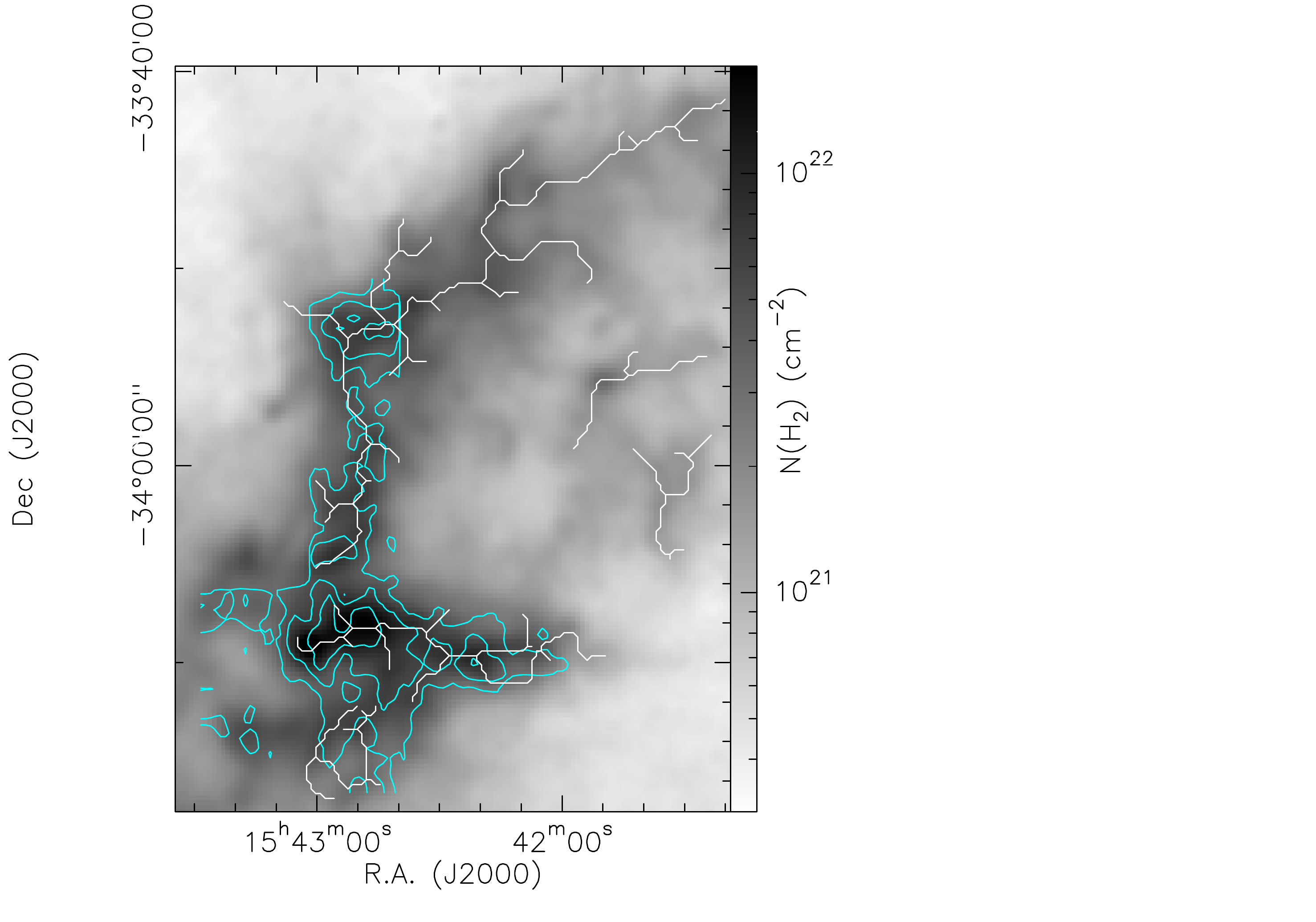}
\caption{A zoom into the densest region of the Lupus 1 cloud that shows the 250 \um\, emission (background), the spine of the identified filaments (white lines) and the CS (2--1) integrated intensity (cyan contours). Note that the CS (2--1) map does not cover the Northern (declination $>$ -33\degr50\arcmin) and the Western (R.A. $<$ 15$^h$42$^m$) part of the map.}
\label{fig_lup1_zoom}       
\end{figure}

\begin{figure}
\includegraphics[scale=.37, angle=90]{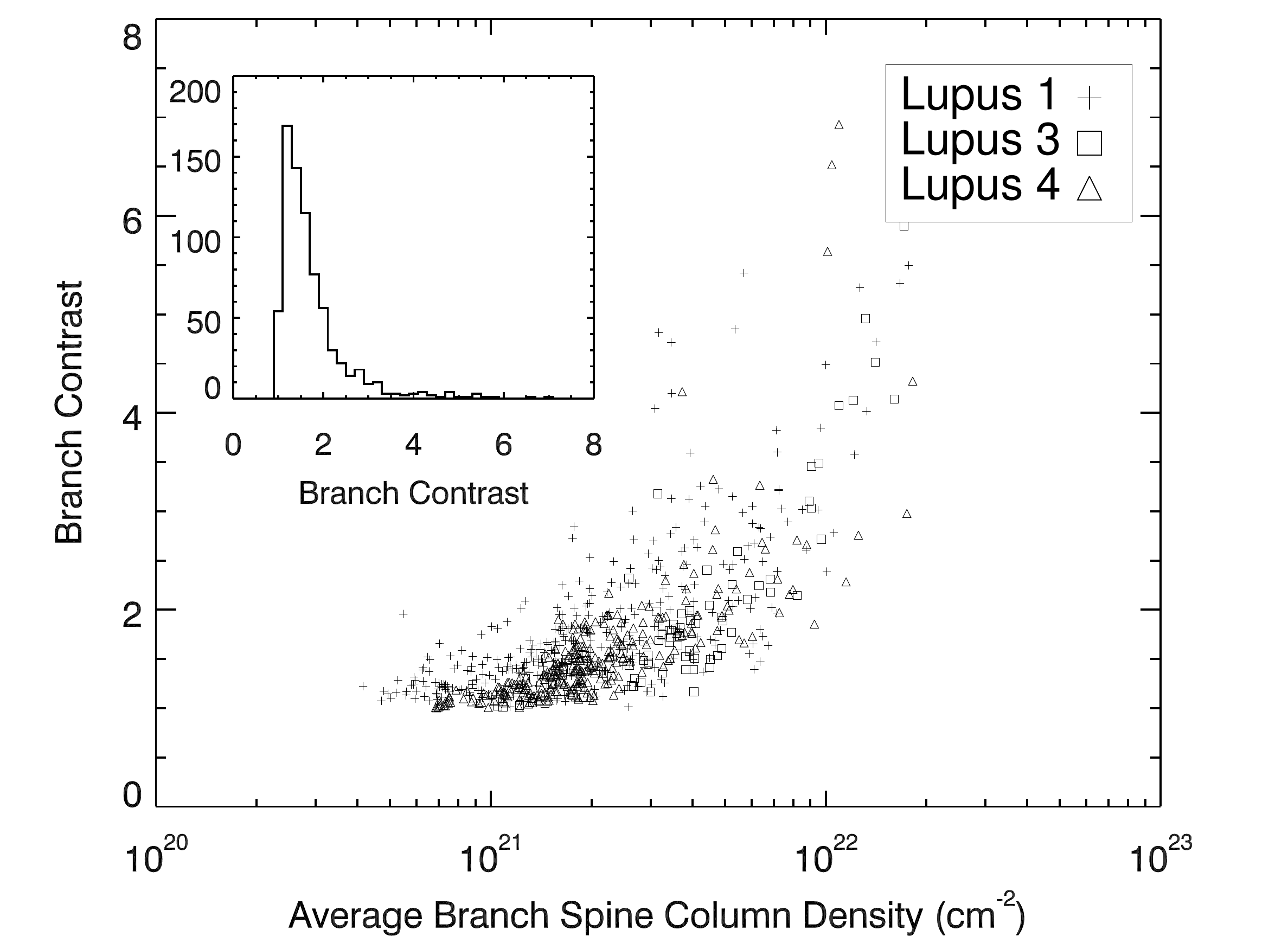}
\caption{Branch contrast, defined as the ratio of the average column density along the spine of the branch and the average column density at the border, {\it vs} the average spine column density. The up-left panel show the histogram of the branch contrast.}
\label{fig_contrast}       
\end{figure}

The filamentary structures identified by the algorithm depend on the adopted threshold on the eigenvalue map. The eigenvalues are a measure of the region contrast with respect to its surrounding, hence the selected threshold determines the lowest contrast that an elongated region should exhibt to be considered  belonging to a filamentary structure. Low threshold values trace low contrast regions, including also several faint structures. We visually inspected the output with different thresholds and choose the value equal to 2.8 time the local standard deviation of the minimum eigenvalue map. In fact such a choise allows to identify the most bright structures and many faint filament branches departing from the bright ones (often called striation). An example is reported in Fig. \ref{fig_lup1_zoom} where the detected filament spines are ovelayed to a portion of the 250 \um\, map of Lupus 1 cloud. As in any thresholding algorithm, we could miss some faint structures. To quantify such effect we measured the branch contrast of the identified filamentary structure defined as the ratio between the average column density along the spine and the average column density at the border.
The histogram of the branches contrast, reported in the nested panel of Fig. \ref{fig_contrast},  shows an increasing numbers of branches for decreasing contrasts up to its peak at contrast around 1.2, testifying that we recover well also the low contrast part of filaments. Furthermore, we report in Fig. \ref{fig_contrast} the branch contrast as function of the average spine column density showing that the denser filaments tend to have the higher contrast but that well defined low density filaments also exist, although our census could to be not complete for branches with very low column density contrast.

Once filaments have been identified, we calculated their physical properties. The width of a filament, deconvolved from the beam, was derived from the median value of the full widths at half maximum (FWHM) of Gaussian fit to the inner part of the column density profile, in the direction perpendicular to the local spine along all the pixels of the spine. The average column density of a filament was calculated by averaging the column density of all the pixels inside the filament border and the mass per unit length of each filament was obtained by multiplying the average column density for the width. We stress that our definition entails global properties for the whole filamentary structure. However, given the complex structure of the longer filaments, we also calculated the same physical parameters at a smaller scale. i.e. for all the branches composing the filaments. In particular, we estimated the direction of the branches from the linear interpolation of the points along the branch central spine.

\begin{figure}
\includegraphics[scale=.48]{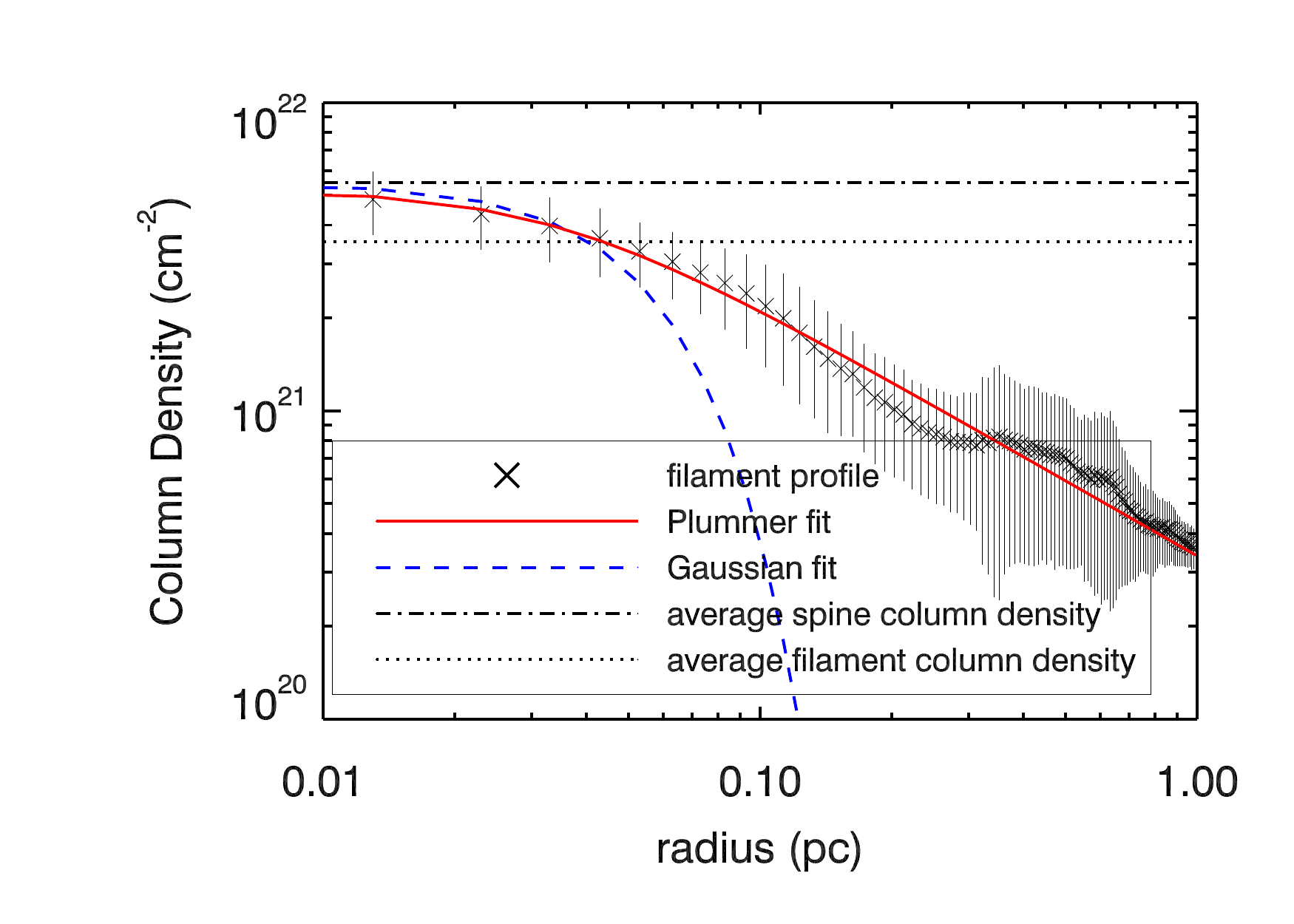}
\caption{Mean radial column density profile (crosses) of the long filament displayed in the Northern part of Fig. \ref{fig_lup1_zoom}. The error bars show the 1$\sigma$ dispersion of the distribution of radial profiles along the brightest part of the filament. The red continuum curve shows the best-fit Plummer model described by Eq. \ref{eq_plummer} with $p$ = 1.8  and $R_{flat}$ = 0.03 pc. The dashed blue curve shows a Gaussian fit to the central part of the profile with a FWHM = 0.07 pc, which is considered as the inner width of the filament. The dot-dashed horizontal line indicates the value of the average column density along the spine while the dotted line indicates the average filament column density.}
\label{fig_fil_profile}       
\end{figure}

An example of the radial column density profile is shown in Fig. \ref{fig_fil_profile}, it is composed by a flat, dense inner region and a power-law behaviour at large radii. This form is typical of filaments detected in the {\it Herschel} maps of low mass molecular clouds (\citealt{arzoumanian11}; \citealt{palmeirim13}). Analytically, it is described by a Plummer-like function of the form
\begin{equation}
 N(\rm{H_2}) =   \frac{N_{c}(\rm{H_2}) R_{flat}}{[1+(r/R_{flat})^2]^{\frac{p-1}{2}}}
 \label{eq_plummer}
\end{equation}
that derives from an idealised model of a cylindrical filament laid on the plane of the sky with radial density of the form 
\begin{equation}
\rho(r) = \frac{\rho_c}{[1+(r/R_{flat})^2]^{p/2}}
\end{equation}
where $\rho_c$ is the central density, $N_{c}(\rm{H_2})$ is the central column density, $R_{flat}$ is the radius of the flat inner region, $p$ is the power-law exponent at large radii. For the filament shown in Fig. \ref{fig_fil_profile} the best fit model has $R_{flat}$ = 0.03 pc and $p$ = 1.8, fully in agreement with values found in other filaments (\citealt{arzoumanian11}; \citealt{palmeirim13}) and significantly shallower than the steep $p$ = 4 profile expected for unmagnetized isothermal filaments in hydrostatic equilibrium \citep{ostriker64}. In Fig. \ref{fig_fil_profile} we also report the average column density along the spine that some authors use as indicative of the  filament column density and the average column density as calculated by us. The two values differ of a factor of about 1.5 well below the uncertainty associated to the filament column density (see Sect. \ref{fil_param}).

\section{filament analysis}

\subsection{Physical properties of the filaments}
\label{fil_param}

In Figs. \ref{fig:1_lup1}, \ref{fig:1_lup3} and \ref{fig:1_lup4} we show the spines of all the branches of the detected filaments on top of the column density maps of the three Lupus clouds. Lupus 1 appears itself as a long ($\sim$ 5.2 pc) filamentary structure extending from north-west to south-east, broken into a sequence of shorter filaments with lengths from 0.1 pc to 2 pc. South-West of the main filamentary structure, a smaller region of dense material exists with a roundish form where other filaments have been identified.
Similar to Lupus 1, the Lupus 3 cloud is composed of two long filamentary structures, running almost parallel in the east-west direction and fragmented into shorter filaments. By contrast, the distribution of filaments in Lupus 4 appears to be more chaotic,  without a clearly defined several pc filamentary structure, even though several long (up to 3 pc) filaments running east-west have been identified.

\begin{figure}
 \includegraphics[scale=.9]{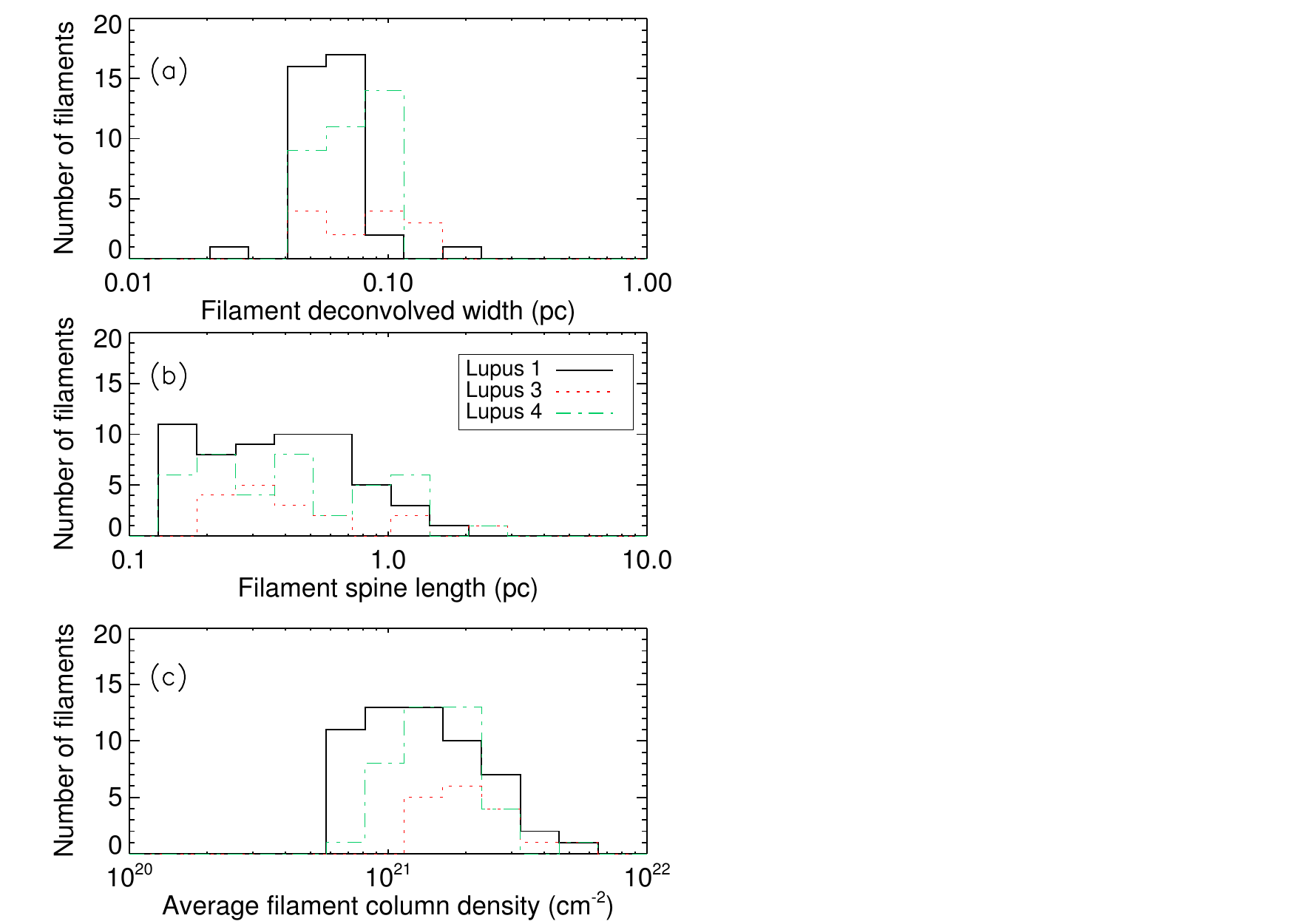}
 \caption{Distribution of the deconvolved width of the filaments (a), the length of the spine of the filaments (b) and the average filament column densities (c) of the three Lupus regions. The spatial resolution limit of the images is 0.026 pc for Lupus 1 and 4 and 0.035 pc for Lupus 3. }
 \label{fil_hist}
\end{figure}

In Fig. \ref{fil_hist}, we present the distributions of the physical properties of the filaments identified in each of the three clouds, namely the deconvolved width, the projected length (defined as the length of the longer spine build with the MST in the filament) not corrected for the inclination angle and the average column density. The median value of the distributions are reported in Table \ref{tab_median}. The median value of the filament width is slightly lower, but still consistent with, the typical value of 0.1 pc found in a wide sample of filaments in other Gould Belt molecular clouds \citep{arzoumanian11}. Filaments are generally short with a median length of $\sim$ 0.4 pc and only a few structures longer than 1 pc. The aspect ratio, however, is quite high, within the range from 3 to 40 with median value of 7, 5 and 6 for Lupus 1, 3 and 4, respectively, indicating the definite filamentary nature of the identified structures. The average filament column density ranges from 6$\times$10$^{20}$ \cmdue\, to 7$\times$10$^{21}$ \cmdue. Noticeably, these values are in the lower part of the distribution of filaments column densities found in other molecular clouds \citep{arzoumanian13}, and they are closer to the column densities found in IC5146  and in the non-star-forming Polaris region rather than to values typical of more active and massive regions such as Aquila in which more massive stars are formed. The derived  average filaments column densities are affected by several sources of uncertainty: {\it i)} the $<$ 20 per cent error of the absolute flux calibration of the original photometric maps; {\it ii)} the uncertainty related to the assumption in the opacity law used to produce the column density maps that ranges from a factor of 1.5 for line of sights toward dense compact cores up to a factor of 2 for low column density regions (see Sect. \ref{sec_cdmap}); {\it iii)} the unknown inclination angle of the filaments that however should be low for filaments having a high aspect ratio, as is the case the ones we have observed in Lupus; and {\it iv)} possible contamination by the background of a more diffuse medium in which filaments could be embedded and that has not been subtracted. In the Lupus clouds, however, the diffuse background is negligible as can be seen in Fig. \ref{fig_fil_profile}, indeed practically all the dense material at visual extinctions larger than 2 mag is arranged in filaments with part of filaments lying in regions of very low column density, at $A_{\rm v}<$ 2 mag, especially in Lupus 1 and 4 (see Figs. \ref{fig:1_lup1}, \ref{fig:1_lup3} and \ref{fig:1_lup4}). Considering the low contribution of the last two points, we estimate an uncertainty on the derived average filaments column density of at most a factor of 2.

In summary, comparing the filaments in the Lupus complex with that of other molecular clouds \citep{arzoumanian13}, and of the Galactic plane \citep{schisano14}, the Lupus filaments have the typical width but a shorter length and, most importantly, a low column density. Approximating the shape of filaments with a cylinder, a fixed observed width and a small column density imply also a small average volume density of the material arranged in filaments. 
In Lupus the low column density regime is not only valid for the material structured in filaments but is a property also of the diffuse material as testified by the probability distribution function analysis (see Sect. \ref{pdf} for a further discussion).

It is worth noting that, in the three clouds both the range and the median values of the analysed filament parameters are very similar (see Fig. \ref{fil_hist}). This result is in agreement with the observed similarity of the overall star formation properties of the three subregions of the Lupus complex. The general uniformity allows us to sum together the filament samples from the three regions, in order to improve the statistical relevance of the subsequent analysis.

\begin{table*}
 \centering
  \caption{Median properties of the filaments distributions in the three Lupus clouds.}
  \begin{tabular}{@{}lccc@{}}
  \hline                                & Lupus 1 & Lupus 3 & Lupus 4 \\
Median deconvolved width (pc)           & 0.05     & 0.09     & 0.07 \\
Median length of the main spine (pc)    & 0.37    & 0.31    & 0.45 \\
Median average column density (10$^{21}$ \cmdue)  & 1.2     & 1.9     & 1.6 \\
Median aspect ratio                     & 7.4     & 4.9     & 5.9 \\
\hline 
\end{tabular}
\label{tab_median}
\end{table*}

\subsection{Filament kinematics}
\label{sect_kin}

\begin{figure*}
\includegraphics[scale=.58]{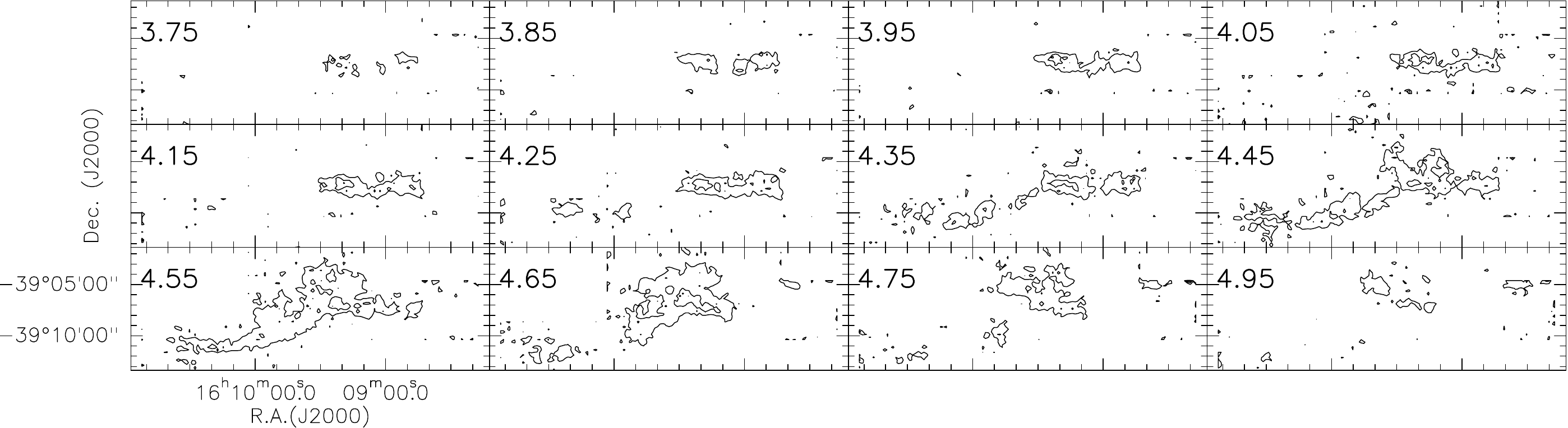}
\caption{CS (2-1) channel map of Lupus 3; first contour and contour steps are 0.1 K \kms.}
\label{cslup3}       
\end{figure*}

Information on the kinematics of filaments can be derived from spectroscopic observations of molecular lines. A good tracer of dense molecular gas is the CS molecule (critical density $\sim$10$^5$ \cmtre).  Accordingly, we used the CS (2--1) line maps of Lupus 1 and 3 for a kinematical analysis of filaments; we do not have data for Lupus 4. The CS (2--1) line maps cover only the denser part of the Lupus 1 and 3 clouds, approximately the area above the level of 4 mag of visual extinction. In the overlapping region the line emission, tracing the molecular gas, shows a morphological distribution which is very similar to the distribution of the dust inferred from the  {\it Herschel} column density map. In fact, the filamentary structures identified in the {\it Herschel} maps are also present in the CS (2--1) line maps, see Fig. \ref{fig_lup1_zoom} as example.

\begin{figure}
\includegraphics[scale=.4]{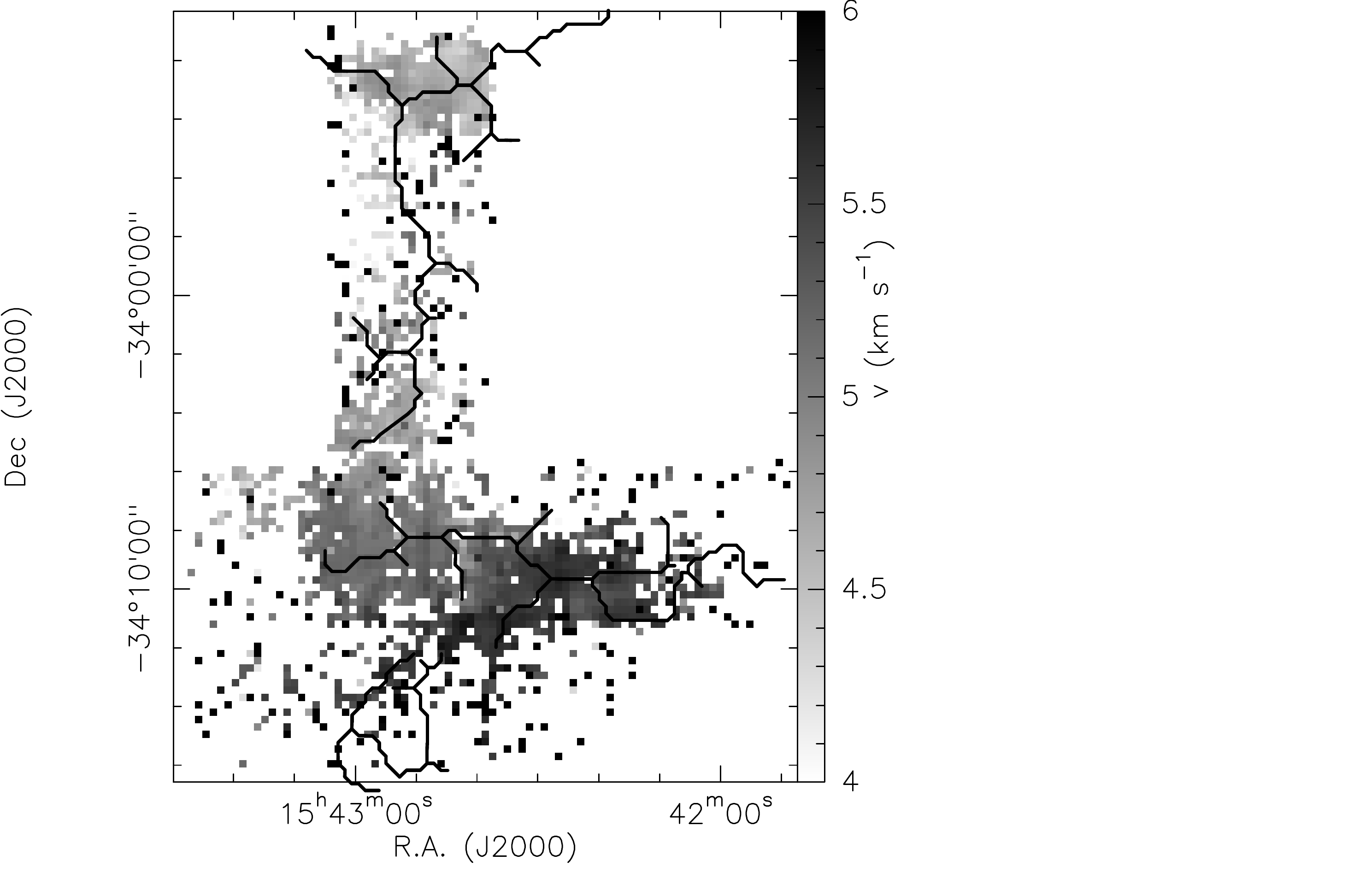}
\caption{Map of the central velocity of the Gaussian fitting of the CS (2-1) line in Lupus 1. The black lines indicate the spine of the filaments identified in the column density map.}
\label{lup1_center_v}     
\end{figure}

\begin{figure}
\includegraphics[scale=.33]{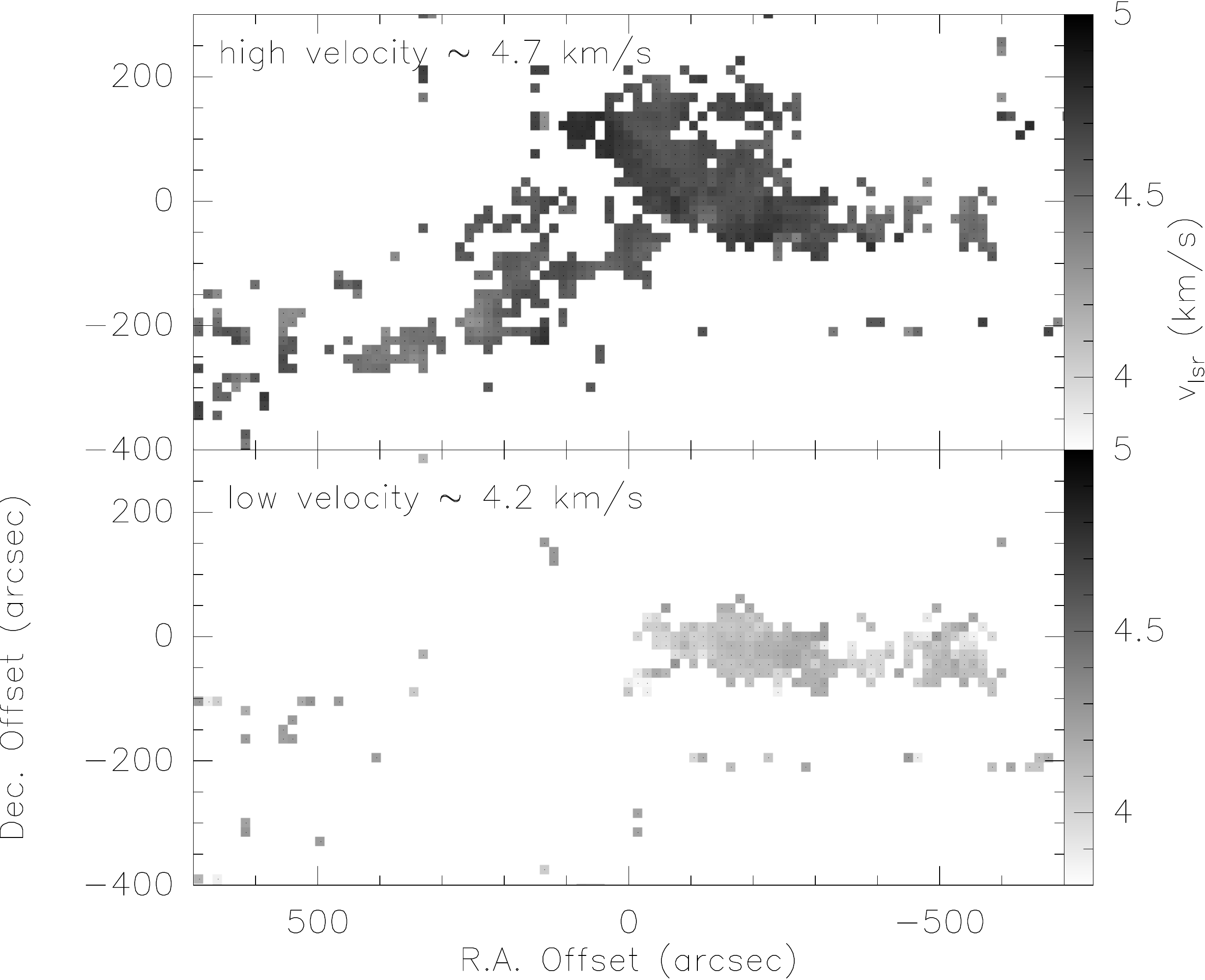}
\caption{Maps of the line centre derived from the Gaussian fitting of the CS (2--1) transition with two components in Lupus 3: the upper panel shows the distribution of the component at higher velocity ($\sim$ 4.7 \kms), the lower panel shows the distribution of the component at lower velocity ($\sim$ 4.2 \kms). In the central part of the map the two components coexist. The spectral resolution of the data is 0.1 \kms. The offset 0,0 corresponds to R.A.(J2000) = 16$^{\rm h}$09$^{\rm m}$34\fs76 and Dec(J2000) = -39\degr06\arcmin54\farcs1.} 
\label{lup3_cs_peak}       
\end{figure}

The CS (2--1) channel maps of Lupus 1 and 3 (see fig. 3 in \citet{benedettini12} and Fig. \ref{cslup3}) show that a single filament or a branch of a filament in the column density maps corresponds to a coherent kinematical structure in the channel maps. However, filaments of one cloud are not all at the same systemic velocity but they are at slightly different $v_{\rm lsr}$, for example at 4.2 \kms\, and 4.7  \kms\, in Lupus 3 and at 4.6 \kms, 5.1 \kms, and 5.8 \kms\, in Lupus 1 (see Figs. \ref{lup1_center_v} and \ref{lup3_cs_peak}).  These differences, well above the spectral resolutionof the data of 0.1 \kms, can be due to different inclination angles of the filaments or/and to small differences in the intrinsic velocity of the material composing the filaments, possibly derived from the turbulence of the original molecular cloud.  The kinematical coherence of filaments and the spread of 1--2 \kms\, of the velocity centroid among different filaments of the same cloud have been observed also in other low-mass star forming regions (e.g. \citealt{hacar11}; \citealt{hacar13}).

From the analysis of the line profile we found that toward the central bright region of Lupus 3, the CS (2--1) line presents two peaks. This can be due to a self-absorption or to the overlap of two different gas components. We do not have the C$^{34}$S (2--1) spectrum to verify which is the true case, however the fact that the peak of lines of different species, namely HC$_3$N (10-9) and N$_2$H$^+$ (1-0) with very similar critical density ($\sim$6$\times$10$^5$ \cmtre) is coincident with only one of the CS peak \citep{benedettini12} supports the second hypothesis. In Fig.~\ref{lup3_cs_peak} we show the map of the line centre of the two components Gaussian fitting: the upper panel shows the distribution of the component at higher velocity ($\sim$ 4.7 \kms), the lower panel shows the distribution of the component at lower velocity ($\sim$ 4.2 \kms), highlighting two coherent structures of gas that superimpose each other in the central part of the map. A similar structure has also been observed in the map of C$^{18}$O (2--1) line obtained with $APEX$ (Benedettini et al., in preparation).

No evidence of velocity gradients in the longitudinal or radial directions of the filaments is revealed in our spectroscopic data, indicating that any possible filament contraction, if present, must produce a velocity gradient, projected along the line of sight, below the  0.1 \kms\, spectral resolution of our data. Given the low value of this upper limit, we can confidently say that there is no evidence of contraction in the Lupus filaments at the present epoch.

\subsection{Filament arrangements}
\label{arangment}

\begin{figure}
 \includegraphics[scale=.5]{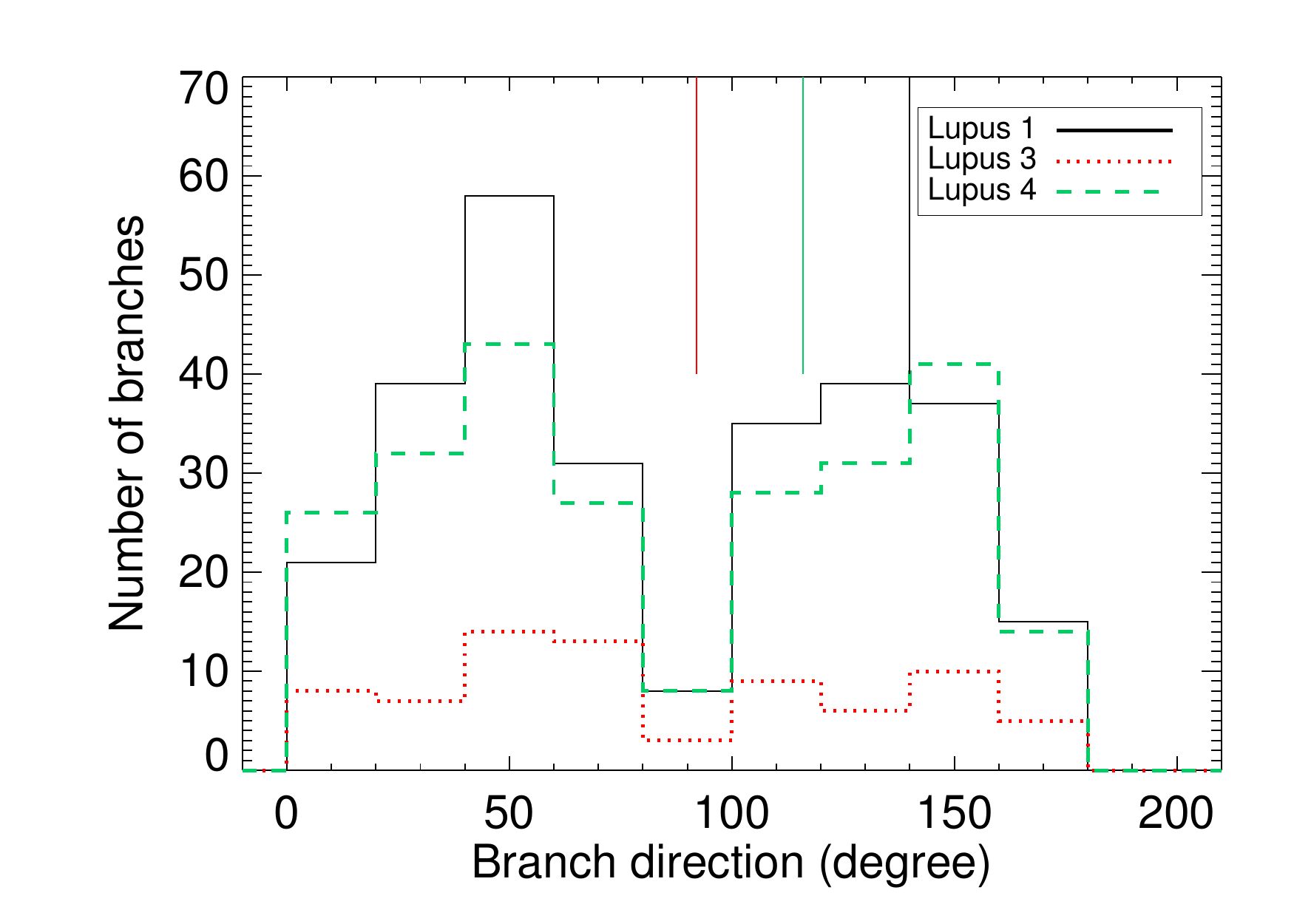}
 \caption{Distributions of the position angles (measured North from West) of the branches of the filaments of the three Lupus regions. The vertical bars indicate the orientation of the ambient magnetic field in each cloud, measured form the polarisation of the light from field stars.}
 \label{fil_angle}
\end{figure}

Looking at the arrangement of the filaments in the Lupus clouds (see Figs. \ref{fig:1_lup1}, \ref{fig:1_lup3} and \ref{fig:1_lup4}), it appears that they are predominantly arranged in the directions of the very long (a few parsec scale) principal filamentary structures. To verify this visual impression, we estimated the direction of the branches of the filaments as the position angle of the linear interpolation of the pixels located on their central spines. The distribution of the position angles of the branches, measured counterclockwise from the west direction (north of west), for the three clouds is shown in Fig. \ref{fil_angle}. The distributions in each Lupus clouds are bimodal with two orthogonal peaks, one around 50\degr\, and the other around 140\degr. The bimodal distribution is less evident in Lupus 3 which can be due to the poor statistics of this cloud. The similarity of the position angle distributions among the three clouds suggests that a common, probably external, event acted in shaping the material of the clouds and the formation of filaments. This common event could be the supernova explosion in the Upper-Sco sub-group of the nearby Scorpius-Centaurus OB 2 association \citep{humphreys78} that has been suggested to be the event that triggered the recent star formation in the Lupus complex \citep{tachihara01}.

Interestingly, one of the peaks in the position angle distribution is similar to the orientations of the magnetic field in the regions (estimated from polarimetric observations of the field stars) that are: 124\degr--140\degr, 92\degr\, and 115\degr\, for Lupus 1, 3 and 4, respectively (\citealt{myers91}; \citealt{rizzo98}; \citealt{matthews14}). Assuming that the magnetic field inside the cloud maintains a similar direction of that of the ambient field, and considering the large dispersion associated to the mean value of the orientation of the magnetic field, we can assert that filaments in Lupus are mainly arranged along the direction of the magnetic field or in the orthogonal direction. This similarity is an indication that the magnetic field may have also had a role in the filaments formation and may be important to determine their support, as also found in other star-forming regions (e.g., \citealt{palmeirim13}; \citealt{busquet13}). Recently \citet{li13} investigated the relation between the magnetic field and the ISM distribution, and found that the few parsec length filamentary structures present in some of the Gould Belt molecular clouds preferentially lay either parallel, or orthogonal to the field direction. Our result in Lupus suggests that this is not only valid for the long filamentary structures, from a few to 10 pc scale investigated by \citet{li13}, but also at the sub-parsec scale probed by our filaments that have length smaller than 1 pc.

\subsection{Association with compact sources}

Filaments have been demonstrated to be the preferred place for the formation of new stars (e.g. \citealt{andre10}; \citealt{polychroni13}; \citealt{konyves15}). However the underlying physical processes that describe how the diffuse gas in the filaments fragments, and condenses into dense cores, remain unclear. The far-infrared {\it Herschel} maps are a powerful dataset for this kind of study since in these maps both filaments and compact dense cores from which stars form can be easily identified. In particular, the {\it Herschel} compact sources represent dusty dense cores that can have different natures, indeed they can be young stars still embedded in their parental envelope, cores without an internal source but dense enough to eventually form a star, or unbound cores that may eventually dissipate. The presence of a protostar that internally accretes from the dense core can be revealed by emission at wavelengths shorter than 100 \um\, that deviates from the grey-body spectral energy distribution typical of the emission at longer wavelengths. Therefore, a detection in the PACS 70 \um\, map can be used to identify the protostars. The further distinction among the starless cores of the gravitationally bound structures that are potential sites of star formation requires the estimate of the core mass. However, a detailed catalogue of {\it Herschel} compact sources is out of the scope of the present paper and it will be the matter of a forthcoming paper (Benedettini et al., in preparation). Here, we refer to the preliminary list of {\it Herschel} compact sources published by \citet{rygl13}. 

Looking at the distribution of those sources (see Fig. A.1 of \citealt{rygl13}) it is clear that the majority of the starless cores (both gravitationally bound and unbound cores) are associated to  filaments, as observed in other star forming regions (e.g. \citealt{andre10}; \citealt{polychroni13}; \citealt{konyves15}). In particular, cores are usually close to the central higher column density regions of the filament (within one beam from the spine). Remarkably, in Lupus we find also filaments, or branches of long filaments, that do {\it not} contain compact sources. Filaments without cores are also observed in Polaris, a quiescent non-star-forming region, while they are uncommon in more active and massive regions as Orion L1641. In Lupus those filaments without dense cores are mostly located in the lower column density regions of the cloud, indicating that the density fluctuations produced by the filament fragmentation may survive and grow preferentially in higher column density filaments.

\section{Probability Distribution Functions}
\label{pdf}

\begin{figure}
\includegraphics[scale=.45]{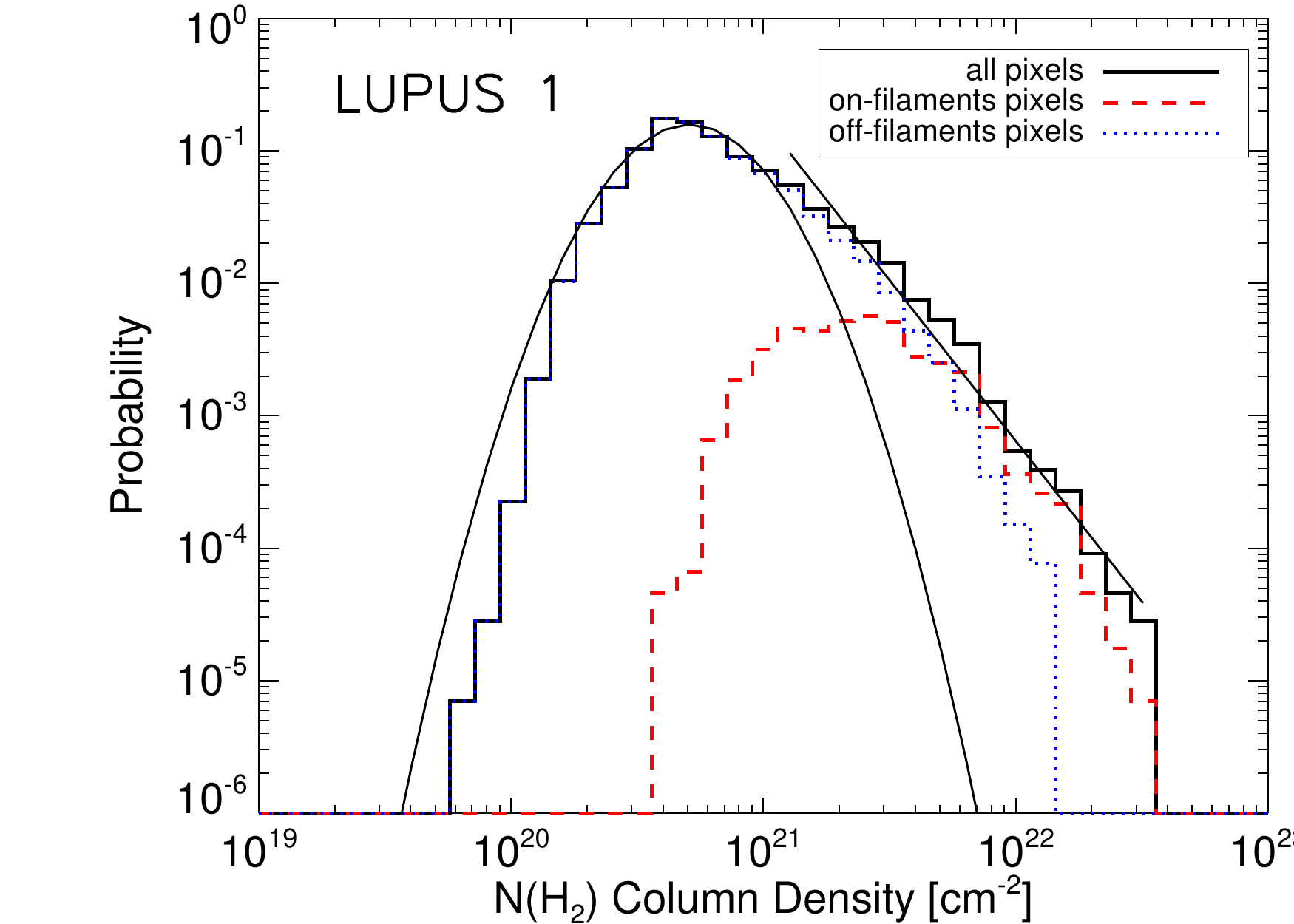}
\includegraphics[scale=.45]{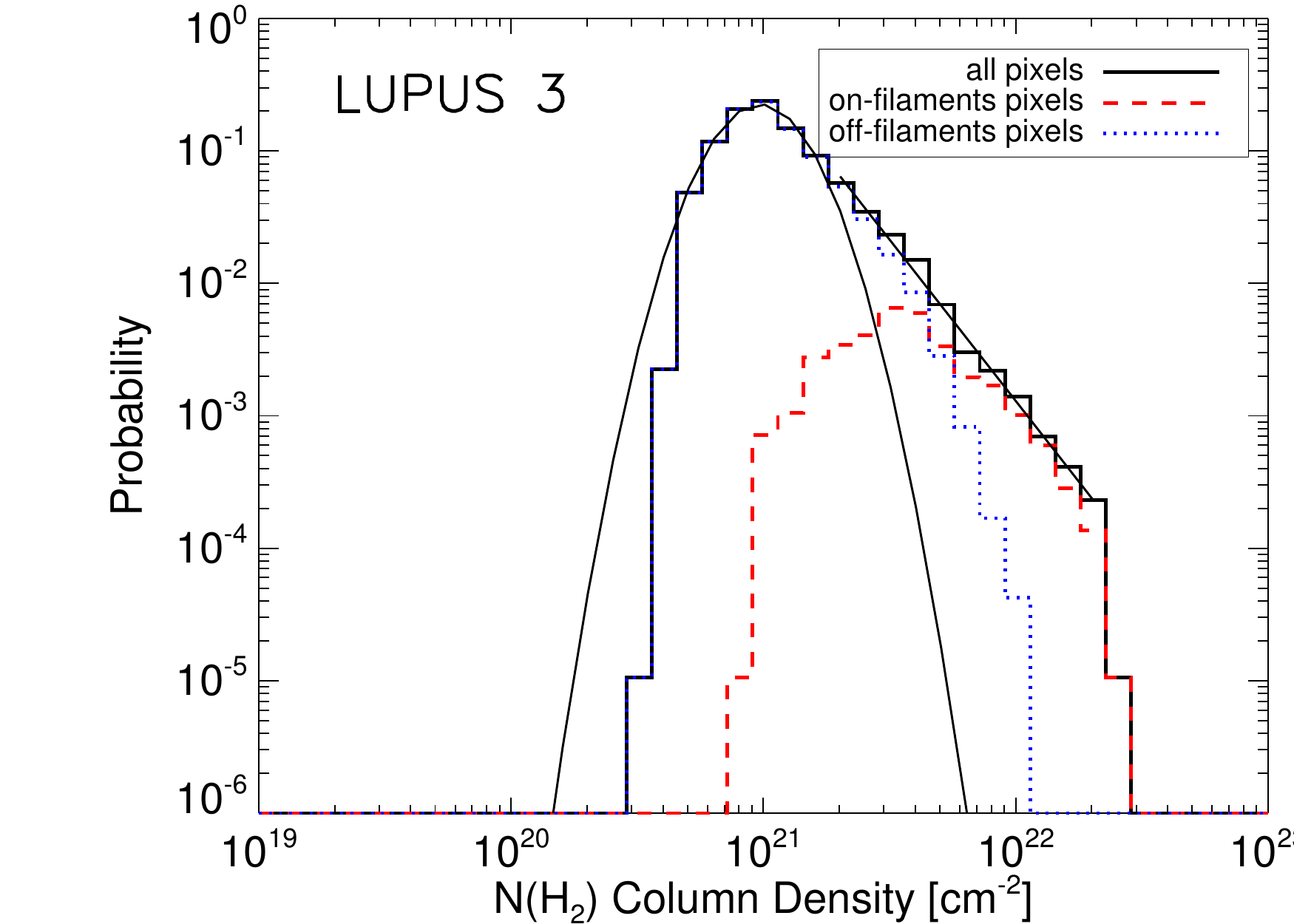}
\includegraphics[scale=.45]{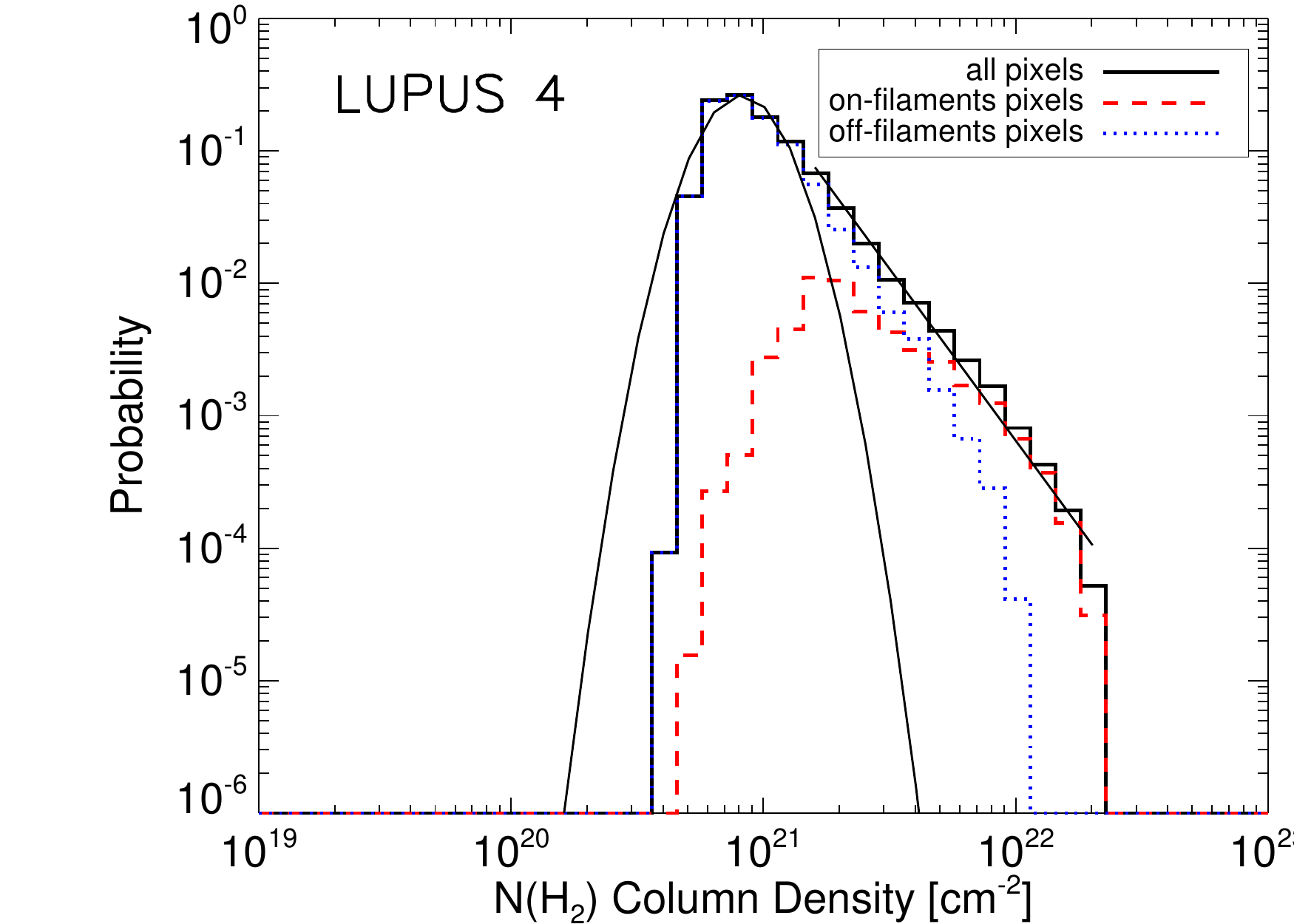}
\caption{PDF of the Lupus 1, 3 and 4 clouds. The black histogram represents the distribution of all the pixels of the map, the red, dashed histogram the distribution of the pixels belonging to filaments excluding pixels around dense cores and the blue dotted histogram the distribution of the pixels not belonging to filaments excluding pixels around dense cores. The log-normal and power law fit to the PDFs are also shown as continuous black lines.}
\label{pdf_fig}       
\end{figure}

The probability distribution functions of column density (PDFs) for Lupus 1, 3, and 4 are displayed in Fig. \ref{pdf_fig}. Their shape is best fitted by a log-normal distribution at low column densities and a power law at higher column densities, confirming earlier studies of PDFs in molecular clouds from extinction maps and {\it Herschel} column density maps (e.g. \citealt{kainulainen09}; \citealt{schneider13}; \citealt{tremblin14}). However, a single log-normal fit for the very low column density range shows a deficit in the observed PDF which may account for a more complicated PDF structure. In particular, noise, definition of boundaries,  image clipping, and projection effects make this low column density end difficult to constrain by observations (\citealt{schneider15}; \citealt{lombardi15}). In Table 3 we report the parameters of the best fitting for the three Lupus clouds. Both the widths of the log-normal part and the slopes of the power-law tail are in agreement with what is found in other {\it Herschel} studies of low- and high-mass star forming clouds (\citealt{hill11}; \citealt{schneider13}, 2015; \citealt{alves14}; \citealt{tremblin14}; \citealt{konyves15}).

Comparing the PDF of the three clouds, it is evident that the PDF of Lupus 1 extends to much lower column density, and has a peak at a value lower of about a factor of two compared to the other two clouds. This characteristic is partially due to the fact that in Lupus 1 we mapped not only the denser part of the cloud but also a large area of its more diffuse surrounding medium. Indeed, about the 75 per cent of the field is composed of pixels at very low column density ($A_v<$ 1 mag), while it is not the case for the other two clouds. The very low column density of the diffuse material around Lupus 1 can be related to its peculiar position. This cloud is at the edge of an HI shell left over from a supernova event in Upper--Scorpius. The dynamics of HI and CO toward Lupus 1 are consistent with the molecular cloud being compressed by the HI shell (\citealt{degeus92}; \citealt{tothill09}). This scenario is supported by the morphology of the dense gas shown by the {\it Herschel} column density map, with the gas at $A_v>$ 1 mag confined in a thin layer parallel to the edge of the HI shell (see Fig. \ref{fig:1_lup1}) surrounded by the very low density medium swept by the wind of the supernova.
On the other hand, Lupus 3 and 4 are further from the ridge of the shell and their dense material is more distributed. The cloud morphology revealed by the {\it Herschel} column density maps  once again indicates the role of the Scorpius--Centaurus OB association in the star formation history of the Lupus complex. 

A peculiarity of the PDFs of the Lupus clouds is that the values of the peak and the maximum of the column densities are systematically lower than those found in active regions such as Orion and Aquila, and appear to be much more similar to the values found in Chamaeleon and Polaris (\citealt{schneider13}; \citealt{schneider15}; \citealt{alves14}), regions with low or null star formation activity. Remarkably, in the Lupus clouds the peak of the PDFs is at H$_2$ column density equivalent to visual extinction less than 1 mag, the lowest value found so far in the PDF analysis of column density maps of star forming regions based on {\it Herschel} data, that usually peak at visual extinction larger than 1.5 mag.
These low values of the H$_2$ column density is an indication that in Lupus a part of the Hydrogen could be still atomic rather than molecular. Indeed, in Lupus the ratio of the mean column density of atomic Hydrogen \citep{degeus92} over the peak the molecular Hydrogen distribution is larger than one ($\sim$ 1.6 and 1.3 in Lupus 1 and 3, respectively) while, for example, in Orion B it is lower $\sim$ 0.6 (\citealt{schneider13}; \citealt{chromey89}).

\begin{table}
 \centering
  \caption{Parameters of the PDF fitting. For the log-normal function: the column density at the peak \textless $N$\textgreater, the width $\sigma$ and the maximum $M$; for the power law tail: the exponent $s$.}
  \begin{tabular}{@{}lccc@{}}
  \hline  
                                     & Lupus 1   	    & Lupus 3 		  & Lupus 4 \\
\textless $N({\rm H_2})$\textgreater (cm$^{-2}$)  & 5.1$\times$10$^{20}$ &9.7$\times$10$^{20}$ & 8.2$\times$10$^{20}$ \\
$\sigma$                             & 0.23       & 0.17		  & 0.14 \\
$M$				     & 0.16       & 0.23		  & 0.26 \\
$s$				     & -2.4	  & -2.4		  & -2.6 \\
\hline 
\end{tabular}
\label{pdfparam}
\end{table}

\begin{figure}
\includegraphics[scale=.43]{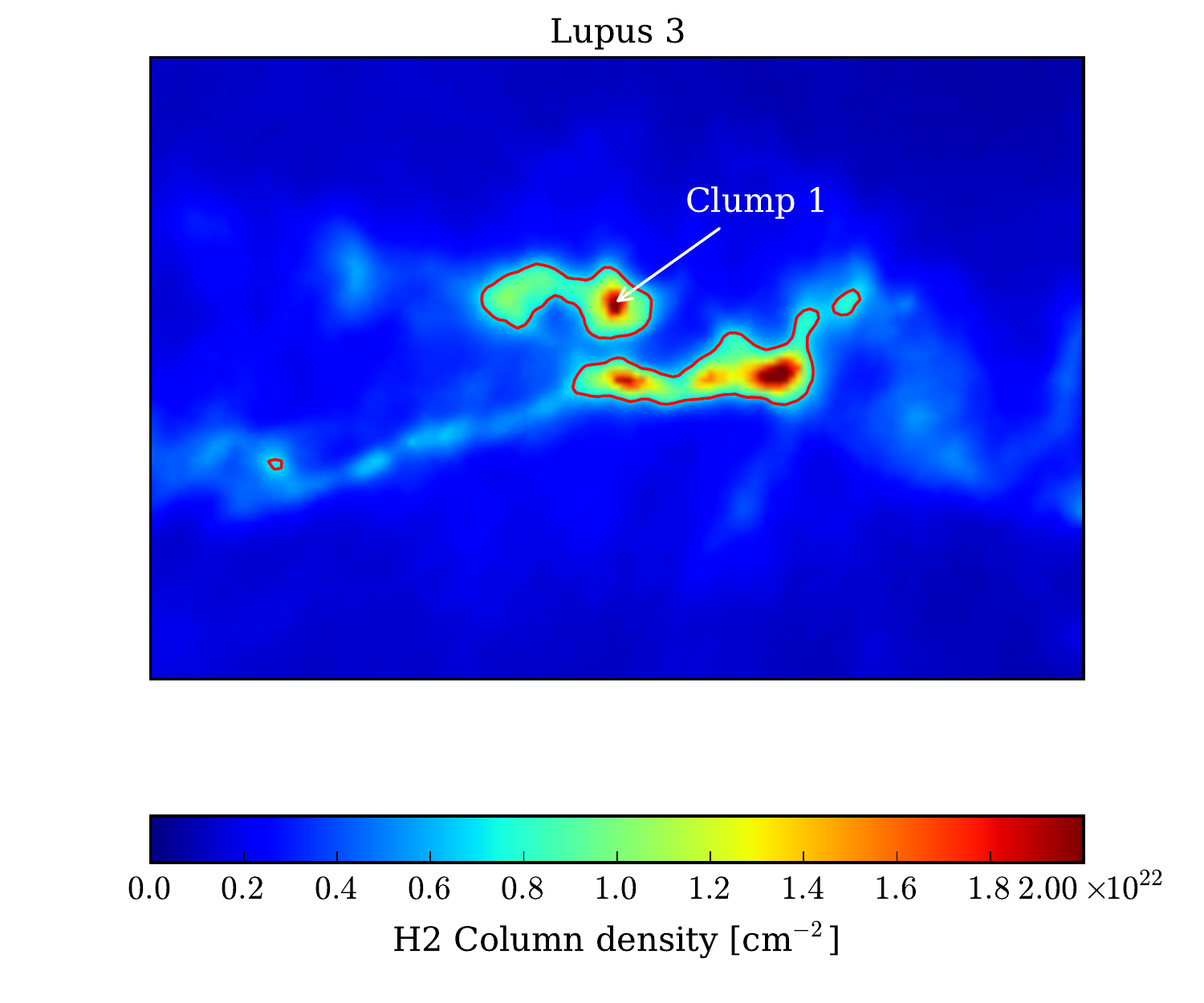}
\includegraphics[scale=.45]{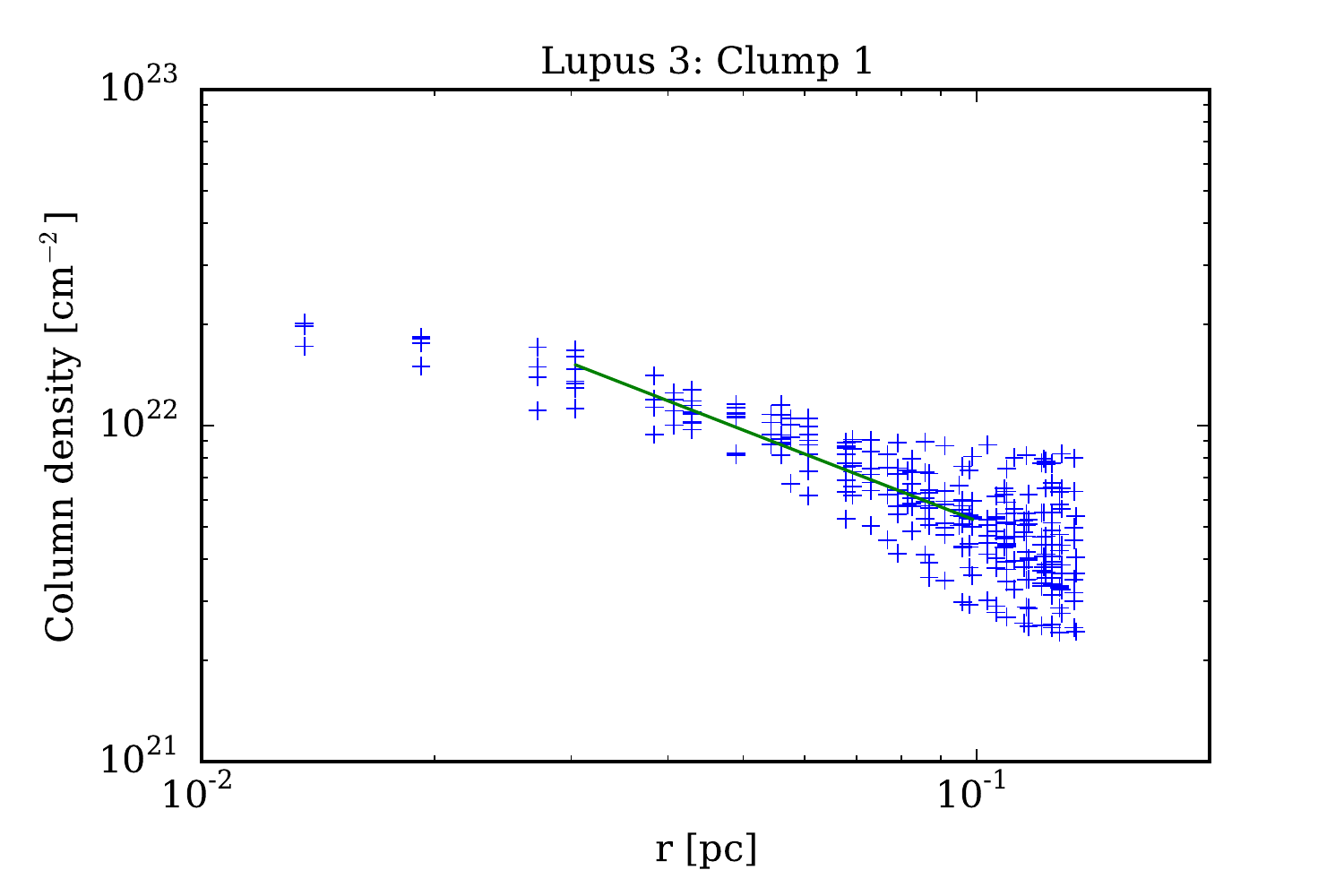}
\caption{{\it Top:} A zoom of the column density map of the central region of Lupus 3. The contour of 7$\times$10$^{21}$ \cmdue\, is drawn, showing that the higher column density material is confined in rather clumpy structures. {\it Bottom:} radial profile of clump 1. We fitted the external part with a function of the form N(H$_2$) $\propto r^{-\alpha}$, obtaining a best-fit value of $\alpha$= 1.9 indicative of a gravitationally bound core.}
\label{fig_clump_lup3}       
\end{figure}

We separated the PDFs into ones with pixels on filaments and one with pixels off filaments (Fig. \ref{pdf_fig}). In this analysis we excluded pixels specifically belonging to dense cores, masking all the pixels within the HPBW from their reference positions. Such exclusion allows us to separate the contribution of the cores from that of the filaments at the high column density end of the distribution, where we expect a strong contribution from gravitationally bound structures. The off-filament PDF shows a log-normal distribution at low column densities but still has a power-law tail at higher column densities, albeit steeper than the total PDF. On the other hand, on-filament PDF makes up the bluk of the power-law tail at higher column densities, confirming that the higher density medium is almost confined in filaments. A similar result has also been found by \citet{schisano14} for filaments in a portion of the Galactic Plane and by \citet{konyves15} in Acquila. Since deviations from the log-normal distribution in the form of a power law at high column densities are predicted for self-gravitating clouds (\citealt{klessen00}; \citealt{ballesteros11}), what we observe is an indication that self-gravity is important in some non-cores portions of the filaments and that locally the denser material contained in filaments, but still not in confined in a core, may undergo a gravitational collapse. This hypothesis can be also supported by the morphology of the material that composes the power-law tail of the PDFs. These pixels can be spatially identified by plotting the contour of the column density at which the on--filament distribution deviates from the off--filaments one. This column density threshold is around 7$\times$10$^{21}$ \cmdue. The structures showed by this contour, although sitted on filaments, are rather clumpy (see Fig. \ref{fig_clump_lup3}) indicating that they are likely affected by gravitational collapse. A further indication of gravitational collapse in the identified structures is the link between their radial profiles and the index of the power--law in the  PDF. A filamentary collapsing structure would give a radial profile $ \propto r^{-\alpha}$ where $\alpha$ is related to the exponential of the PDF power law tail $s$ through the relation $\alpha$=1-1/$s$. Assuming for $s$ the fitted values  given in Table \ref{pdfparam}, the corresponding filament profile would be very shallow with $\alpha$=1.38, 1.40 and 1.42 for Lupus 1, 3 and 4, respectively. By contrast, clumpy collapsing regions would have a radial profile $r^{-\alpha}$ with $\alpha$=1-2/$s$, namely $\alpha$=1.77, 1.80 and 1.83 for Lupus 1, 3 and 4, respectively. These profiles fit very well with the classical values given from gravitationally-bound (free--fall and collapsing) structures with $\alpha$ between 1.5 and 2. Furthermore, we checked the radial profiles of the regions above the 7$\times$10$^{21}$ \cmdue\, threshold in Lupus 1 and 3 (an example is shown in  Fig. \ref{fig_clump_lup3}) and we found $<\alpha>$=1.89 for Lupus 1 and $<\alpha>$=1.95 for Lupus 3, indicating local and clumpy collapsing regions rather than filamentary regions.

\section{Discussion}
\label{discussion}

\subsection{Lupus: a prototype of low column density star forming region}
Analysing the morphological distribution of the dense material in the Lupus complex, we find that almost all the material at visual extinctions larger than 2 mag is arranged in filaments and that some filamentary structures are present also in less dense region, e.g. in zones at $A_{\rm v} <$ 1 mag.
The Lupus clouds are characterised by having low column densities, both in the dense material arranged in filaments, whose average column density is $\sim$1.5$\times$10$^{21}$ \cmdue, as well as in the diffuse medium, as seen from the peaks in the PDF distributions between 5$\times$10$^{20}$\cmdue\, and 10$^{21}$\cmdue. The absence of very high column density gas and the predominance of low column density gas of course have an impact on the typical mass of the stars formed in Lupus that is quite low. Lupus is in fact one of the main low-mass star-forming complexes and its stellar population is dominated by mid M-type stars \citep{hughes94} with a relative fraction of low-mass stars greater than other low-mass star-forming regions such as Taurus. In comparison to other molecular clouds, the star formation regime in Lupus, in terms of star formation rate and stellar clustering, represents an intermediate case between heavily clustered, active and more isolated, quiescent star forming regions. 
Our study reveals that also the diffuse matter from which the stars are formed shows the same property: both the PDFs and the distribution of the filaments column density in Lupus are more similar to the values found in the quiescent Polaris region, rather than in the more active and massive Aquila and Orion star-forming regions. 

\subsection{Can thermally subcritical filaments contain prestellar cores?}
\begin{figure}

 \includegraphics[scale=.48]{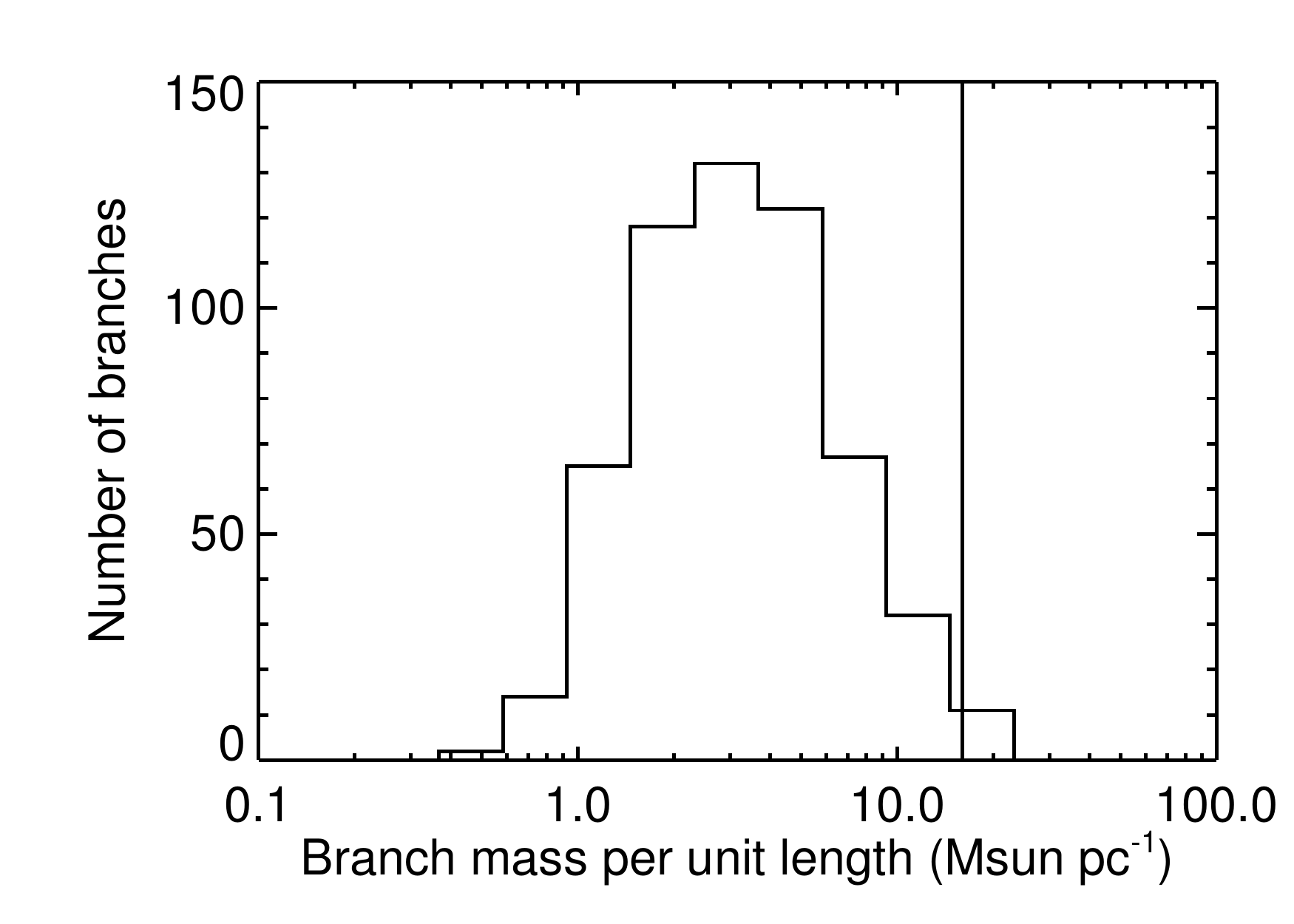}
 \caption{Distribution of the average mass per unit length of the filament branches.The vertical bar indicates the critical value of 16 \msun\, pc$^{-1}$.}
 %containing cores (continuous line) and branches without cores (dashed line) rescaled at the same maximum value. The vertical bars show the median of the two distributions. The red vertical bar indicates the critical value of 16 \msun\, pc$^{-1}$.}
 \label{fig_linemass}
\end{figure}

A key physical parameter for filaments is the mass per unit length. Theoretical models of isothermal infinite cylindrical filaments confined by the external pressure of the ambient medium, predict the existence of a maximum equilibrium value to the mass per unit length \citep{ostriker64}:
\begin{equation}
M_\mathrm{line,max}=\frac{2kT}{\mu m_\mathrm{H}G} = 16.4 \left( \frac{T}{10~{\rm K}}\right) ~ \mathrm{M_\odot~ pc^{-1}}
\label{approx}
\end{equation}
where $\mu$ is the mean molecular weight, $m_\mathrm{H}$ is the mass of the hydrogen atom, $k$ the Boltzmann constant, and $G$ the gravitational constant. When the filament mass per unit length is higher than this maximum value the filament is not thermally supported and it will collapse. In Fig. \ref{fig_linemass}, we show the distribution of the masses per unit length of the branches of the filaments in Lupus. We found that almost all the filament branches have mass per unit length lower than the critical value of 16.4 \msun\, pc$^{-1}$ for radial gravitational collapse and fragmentation of a cylindrical, isothermal filament with a typical temperature of 10 K. Previous studies (\citealt{andre10}, 2014; \citealt{arzoumanian13}) have suggested  that only filaments with mass per unit length above this critical value (thermally supercritical filaments) are associated to star formation activity. 
However, in Lupus there is an evident star formation activity even if the large majority of filaments are subcritical. More specifically, we have found that some subcritical filaments contain starless cores that have been classified as gravitationally bound cores by \citet{rygl13}, since they satisfy the condition $M_{\rm core}\geqslant M_{\rm BE}$, where $M_{\rm BE}$ is the Bonnor -- Ebert mass of a core with equivalent size and temperature of the observed core. The gravitational bound cores are considered good prestellar core candidates in which new stars may form. In \citet{rygl13} the mass of the sources have been derived from a grey-body fitting of the SED of the fluxes measured the {\it Herschel} maps with the {\sc cutex} algorithm \citep{molinari11}. Anyway, disentangling the emission of a compact source from the emission of the embedding filament in the {\it Herschel} maps is not a trivial task. The most critical factors are the fit of the complex background and of the source size that introduces large uncertainty in the core mass estimate and therefore in establishing if the core is gravitationally bound or not. For this reason it is of fundamental importance, when estimating the flux of the sources in a complex region like molecular clouds, to compare the results of at least two completely independent methods to have an idea of the robustness of the measured fluxes. 

\begin{figure}
 \includegraphics[scale=.35]{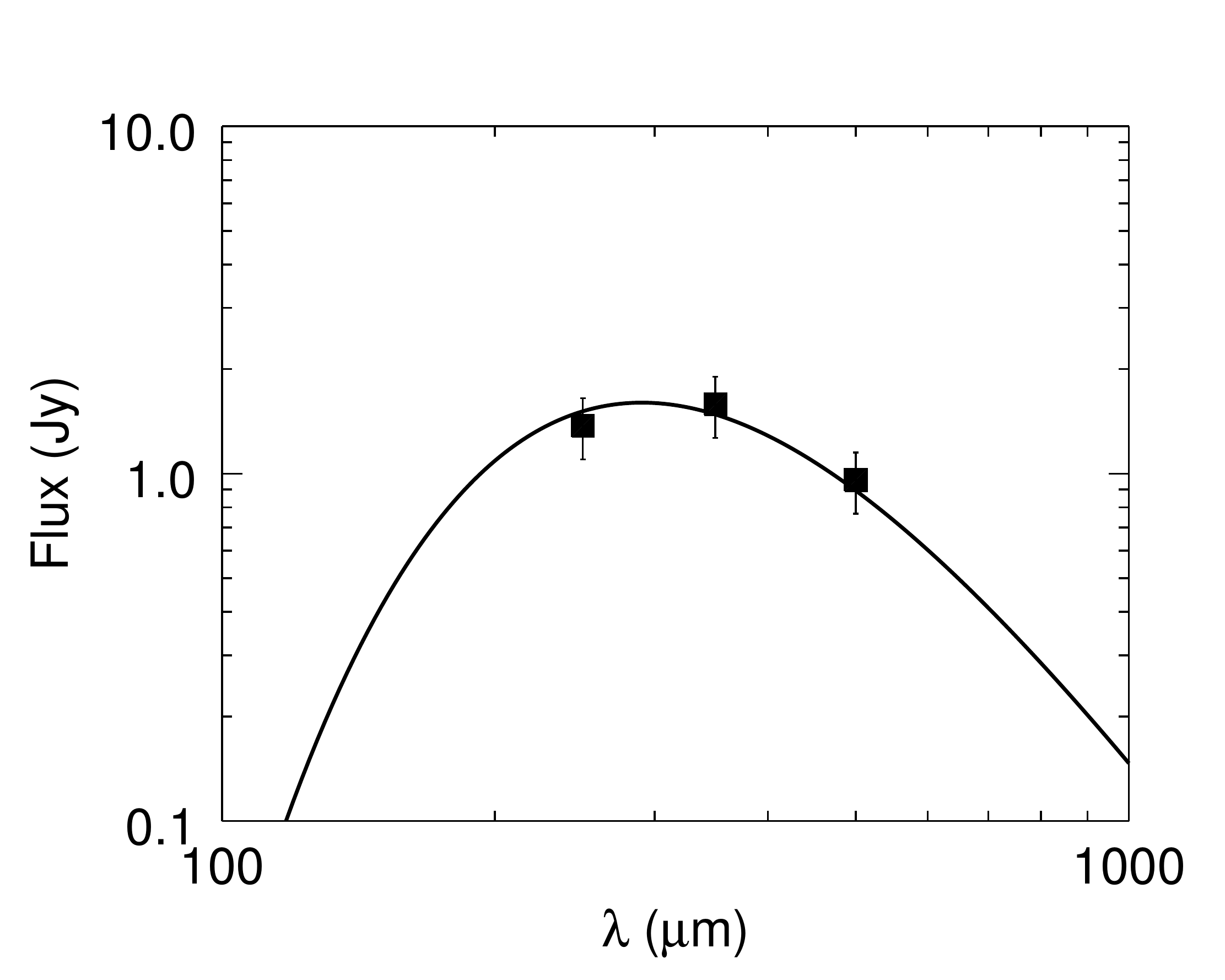}
 \caption{SED of the L1-11 core and grey--body fitting with $T$ = 10 K and $\beta$=2. The error associated to the fluxes corresponds to 20 per cent. See Fig. A.2 in \citet{rygl13} for the five {\it Herschel} maps of the L1-11 core.}
 \label{SEDs}
\end{figure}

To this aim we selected one well defined compact source in Lupus 1 -- the prestellar core L1-11 in \citet{rygl13} at R.A.(J2000) = 15$^{\rm h}$42$^{\rm m}$50\fs6 and Dec(J2000) = -34\degr13\arcmin25\farcs4 -- and we extracted the fluxes at 250 \um, 350 \um\, and 500 \um\, in our new {\it Herschel} maps by using the {\sc hyper} algorithm \citep{traficante15} which computes aperture photometry instead of synthetic photometry like 
{\sc cutex}.
The corresponding SED is reported in Fig. \ref{SEDs} together with a grey-body fitting with fixed $T$ = 10 K and $\beta$ = 2. The best fit core mass is 0.25 \msun, similar to the Bonnor -- Ebert mass of a core with the same size of 36 arcsec and temperature of 10 K, $M_{\rm BE}$ = 0.26 \msun.
In conclusion, the {\sc hyper} photometry confirms the result previously found by \cite{rygl13} by using the {\sc cutex} photometry that this source is a good candidate prestellar core.

Interestingly, this candidate prestellar core is associated to a subcritical filament with a mass per unit length of 5 \msun\, pc$^{-1}$. Note that the filament can be still classified as subcritical even if we consider the largest uncertainty quoted for the column density, namely a factor of 2.
We stress that our estimate of the filament mass per unit length is a value averaged along the branch. In other words, we are assuming that the total mass measured in the entire structure was initially uniformly distributed along the filament branch when it formed. Therefore, even if the branch is on average subcritical, the presence of a gravitationally bound core is a indication that locally, in a smaller portion of the filament, the density is enough to become gravitationally unstable. The demonstration of if and how much the association of prestellar cores with thermally subcritical filaments is statistically significant requires the complete catalogue of prestellar cores in this region and it is deferred to a forthcoming paper (Benedettini et al., in preparation).

\subsection{The role of the magnetic field}

In Sect. \ref{arangment} we showed that the filaments orientation in Lupus is correlated with the local magnetic field. The action of the magnetic field on filaments has been considered by models of \citet{fiege00} and \citet{fiege04}. In these models, the ratio between the mass per unit length and the virial mass per unit length $M_\mathrm{line,vir}=2\sigma_{tot}^2/G$ is a key parameter of a filament that indicates whether it is in pressure equilibrium ($M_\mathrm{line}/M_\mathrm{line,vir}\sim$1) or a poloidal ($M_\mathrm{line}/M_\mathrm{line,vir}\gg$1) or toroidal ($M_\mathrm{line}/M_\mathrm{line,vir}\ll$1) field is required to confine the filament. We calculated the virial mass per unit length by using the total velocity dispersion of the particles
\begin{equation}
 \sigma_\mathrm{tot} = \sqrt{\sigma^2_\mathrm{therm}(\mu) +  \sigma_\mathrm{nonth}^2}
\end{equation}
where the thermal dispersion is
\begin{equation}
 \sigma_\mathrm{therm}(\mu)=\sqrt{\frac {kT}{\mu m_\mathrm{H}}}
\end{equation}
and the non-thermal velocity dispersion is given by
\begin{equation}
 \sigma_\mathrm{nonth}=\sqrt{\sigma_\mathrm{obs}^2-\sigma^2_\mathrm{therm}(\mu_\mathrm{obs})}
\end{equation}
where $\mu_\mathrm{obs}$ is the mean molecular weight of the observed molecule.

We derived the non-thermal contribution for the filaments in Lupus from the line width of the CS (2--1) line ($\mu_\mathrm{obs}$=44), assuming that the line is optically thin. The line was observed with the {\it Mopra} telescope \citep{benedettini12}, at a spatial resolution of 35 arcsec, similar to that of our column density maps. The typical FWHM is about 0.8 \kms\, and 0.4 \kms\, for Lupus 1 and 3 respectively that correspond to a total velocity dispersion of 0.39 and 0.25 \kms, values in agreement with the typical values found in thermically subcritical filaments \citep{arzoumanian13}. The  corresponding virial masses per unit length (70 \msun\, pc$^{-1}$ for Lupus 1 and 30 \msun\, pc$^{-1}$ for Lupus 3) are higher than average masses per unit length measured in Lupus indicating that the observed filaments are gravitationally unbound structures. Therefore, it is possible that an external agent will be required to provide confinement of the filaments. Since we find that $M_\mathrm{line}/M_\mathrm{line,vir}\ll$ 1 a binding, toroidal dominated, magnetic field is possibly confining filaments rather than just the external pressure, in agreement with what found also in other filamentary clouds \citep{fiege00}.

How filaments form from the diffuse material of the molecular clouds is still unknown. Models for filament formation invoke shock compression (e.g., \citealt{heitsh08}; \citealt{padoan01}),  shear and energy dissipation in large-scale MHD turbulence flow (e.g., \citealt{hennebelle13}) requiring the presence of strong magnetic fields. The first scenario foresees randomly distributed filaments without any correlation between the directions of clouds and mean magnetic field directions (Padoan et al. 2001). In the second scenario, however, the competition between gravitational and turbulent pressures in a medium dominated by magnetic fields will shape the cloud to be elongated either parallel or perpendicular to the magnetic field when reaching equilibrium. In contrast to the shocked cloud model, streaming motions are not expected in the equilibrium stage of the magnetic field dominated model. In Lupus we observe a significant alignment of the filament directions with the ambient magnetic field and the absence of gas motions inside the filament, both indications of a significant role of the local magnetic field in the formation of filaments.
On the other hand, self-gravity seems not to play a major role in the filaments formation and fragmentation. Indeed, most of the filament branches in Lupus have average masses per unit length smaller than the critical value for radial gravitational collapse and no evidence of contraction has been seen in the gas kinematics.

\subsection{Kinematical structure}

The kinematical analysis of the filaments identified in column density maps of the Lupus molecular clouds showed that they are composed by a network of short, coherent kinematical structures at slightly different velocity. Similarly, the C$^{18}$O (1--0) map of the B211/3 filament in Taurus revealed that the mostly continuous 10 pc long structure \citep{palmeirim13}, visible in the column density map, is actually composed of many distinct components of gas, each of which has a coherent thermal velocity profile and typical length of 0.5 pc \citep{hacar13}. The inspection of the SPIRE 250 \um\, map, where the small scale structure of the dust emission is better defined thanks to the higher spatial resolution and sensitivity, shows and excellent correspondence between the dust continuum B211/3 filament and the velocity-coherent fibres traced in C$^{18}$O \citep{andre14}. The median length of the coherent kinematical filaments observed in Lupus is similar to the typical length of the fibres observed in Taurus. 

This structure of interwined fibres was not observed in intermediate- and high-mass star-forming regions that have filaments usually thermally supercritical and with much higher column densities than observed in Lupus (\citealt{arzoumanian13}; \citealt{hill11}). Moreover, the kinematical analysis of some massive filaments have shown the presence of velocity gradients indicative of contraction and gravitational collapse (\citealt{schneider10}; \citealt{kirk13}; \citealt{peretto14}), not observed in the Lupus low density filaments.

\section{Conclusions}

We have presented column density maps derived from the SED fitting of the fluxes at four far-infrared bands observed by {\it Herschel} for the Lupus 1, 3, and 4 star-forming regions. From these maps we extracted the filaments and calculated their physical properties. 
The Lupus clouds are characterised by low column densities both for the denser material assembled in filaments, whose average column density is $\sim$1.5$\times$10$^{21}$ \cmdue, and from the diffuse medium, as testified from the peak of the PDFs between 5$\times$10$^{20}$ \cmdue\, and 10$^{21}$\cmdue. This low column density regime of course reflects upon on the mass of the formed stars in Lupus that indeed is low. 
%However, the comparison of the extinction maps derived from the {\it Herschel} maps with those derived from star counts indicates that either in this complex part of the gas could be in form of atomic Hydrogen and therefore not well traced by the {\it Herschel} data or the dust opacity is lower than the values usually found in denser region.

Comparing our filaments catalogue with the preliminary list of prestellar and protostellar sources identified by \citet{rygl13} in the {\it Herschel} maps, we find that most of the sources are located in filaments, in particular close to the central densest parts, testifying that filaments are fragmented and contain the seeds for the formation of new stars. However, we find that, if we consider some average physical properties such as the average mass per unit length, most of the filaments in Lupus turn out to be gravitationally unbound, having average mass per unit length lower than their respective virial critical masses. In addition, they seem to be stable self-gravitating structures, having average masses per unit length lower than the maximum critical value for radial gravitational collapse of an isothermal infinitely long filament. Indeed, no evidence of filament contraction has been seen in the gas kinematics and each filament has an independent, coherent velocity structure. 

On the other hand, prestellar cores and protostars are observed in some of these supposedly globally subcritical filaments. This occurrence is an indication that in the low density regime the critical condition for the formation of stars can be reached only locally and it is not a global property of the filament. In fact, we have found evidences of the predominance of self-gravity in the highest column density regions of filaments in the fact that the power law tail of the observed PDFs are composed mostly by the pixels of the filaments and that these pixels belong to roundish clumpy structures with a radial profile compatible with clumpy collapsing regions. 

Multiple observational evidences indicate the key role that the magnetic field plays in the filament formation and confinement in this region: {\it i)} the masses per unit length are lower than the virial critical value,  {\it ii)} one of the two peaks of the distribution of the position angle of the branches corresponds to the direction of the magnetic field, and {\it iii)} the absence of streaming motions observed in the filament kinematics.

\section*{Acknowledgements}
We thank J.P. Bernard for providing us with the maps of the modelled dust emission at the {\it Herschel} wavelengths that we used for removing the Moon stray-light and D. Eden that produced the visual extinction maps. We are grateful to the anonymous referee for his/her comments that helped us to improve the paper. D.E. is supported by an Italian Space Agency (ASI) fellowship under contract number I/005/11/0. N.S. and P.A. acknowledge support by the ANR-11-BS56-010 project “STARFICH”. This work has received support from the European Research Council under the European Union's Seventh Framework Programme (ERC Advanced Grant Agreement no. 291294 - ORISTARS). PACS has been developed by a consortium of institutes led by MPE (Germany) and including UVIE (Austria); KU Leuven, CSL, IMEC (Belgium); CEA, LAM (France); MPIA (Germany); INAF--IFSI/OAA/OAP/OAT, LENS, SISSA (Italy); IAC (Spain). This development has been supported by the funding agencies BMVIT (Austria), ESA--PRODEX (Belgium), CEA/CNES (France), DLR (Germany), ASI/INAF (Italy), and CICYT/MCYT (Spain).
SPIRE has been developed by a consortium of institutes led by Cardiff University (UK) and including Univ. Lethbridge (Canada); NAOC (China); CEA, LAM (France); IFSI, Univ. Padua (Italy); IAC (Spain); Stockholm Observatory (Sweden); Imperial College London, RAL, UCL--MSSL, UKATC, Univ. Sussex (UK); and Caltech, JPL, NHSC, Univ. Colorado (USA). This development has been supported by national funding agencies: CSA (Canada); NAOC (China); CEA, CNES, CNRS (France); ASI (Italy); MCINN (Spain); SNSB (Sweden); STFC, UKSA (UK); and NASA (USA).

%no commas after author surnames, and no ampersand between the final two author names. List all authors if eight or fewer, otherwise first author only followed by `et al.'.Smith A., 2000, in Minh Y.C., van Dishoeck E.F., eds, Proc. IAU Symp. 197, Astrochemistry: from Molecular Clouds to Planetary Systems. Astron. Soc. Pac., San Francisco, p. 210

\appendix

\label{lastpage}

\end{document}